\documentclass[10pt, doublecolumn]{IEEEtran}
\usepackage{graphicx}
\usepackage{epsfig,latexsym}
\usepackage{float}
\usepackage{indentfirst}
\usepackage{amsmath}
\usepackage{bm}
\usepackage{amssymb}
\usepackage{times}
\usepackage{enumitem}
\usepackage{comment}
\usepackage{caption2}
\usepackage[caption=false,font=footnotesize]{subfig}
\usepackage{algorithm}
\usepackage{algorithmic}

\usepackage{graphicx}
\usepackage{psfrag}
\usepackage{hyperref}
\usepackage{cite}
\usepackage{lastpage}
\usepackage{fancyhdr}
\usepackage{color} 
\usepackage{amsthm}
\usepackage{bigints}
\usepackage{array}
\usepackage{booktabs}
\def\max{\textrm{max}}
\def\min{\textrm{min}}
\newtheorem{remark}{Remark}
\newtheorem{theorem}{Theorem}

\newtheorem{lemma}{Lemma}

\newtheorem{corollary}{Corollary}

\makeatletter

\begin{document}
\title{Exploiting Pinching-Antenna Systems \\ in Multicast Communications}
\author{Shan Shan,~\IEEEmembership{Graduate Student Member, IEEE}, Chongjun Ouyang,~\IEEEmembership{Member, IEEE}, Yong Li,~\IEEEmembership{Member, IEEE},\\ and Yuanwei Liu,~\IEEEmembership{Fellow, IEEE}
\thanks{Shan Shan and Yong Li are with the School of Information and Communication Engineering, Beijing University of Posts and Telecommunications, Beijing 100876, China (e-mail: \{shan.shan, liyong\}@bupt.edu.cn). Chongjun Ouyang is with the School of Electronic Engineering and Computer Science, Queen Mary University of London, London E1 4NS, U.K. (e-mail: c.ouyang@qmul.ac.uk). Yuanwei Liu is with the Department of Electrical and Electronic Engineering, The University of Hong Kong, Hong Kong (e-mail: yuanwei@hku.hk).}
}

\IEEEaftertitletext{\vspace{-2.5em}}
\maketitle
\begin{abstract}
The pinching-antenna system (PASS) reconfigures wireless links through pinching beamforming, in which the activated locations of pinching antennas (PAs) along dielectric waveguides are optimized. This article investigates the application of PASS in multicast communication systems, where pinching beamforming is designed to maximize the multicast rate. 
i) In the single-waveguide scenario, a closed-form solution for the optimal activated location is derived under the assumption of a single PA and linearly distributed users. Based on this, a closed-form expression for the achievable multicast rate is obtained and proven to be larger than that of conventional fixed-location antenna systems. For the general multiple-PA case with arbitrary user distributions, an element-wise alternating optimization (AO) algorithm is proposed to design the pinching beamformer. 
ii) In the multiple-waveguide scenario, an AO-based method is developed to jointly optimize the transmit and pinching beamformers. Specifically, the transmit beamformer is updated using a majorization-minimization (MM) framework together with second-order cone programming (SOCP), while the pinching beamformer is optimized via element-wise sequential refinement. 
Numerical results are provided to demonstrate that: i) PASS achieves significantly higher multicast rates than conventional fixed-location antenna systems, particularly when the number of users and spatial coverage increase; ii) increasing the number of PAs further improves the multicast performance of PASS.
\end{abstract}

\begin{IEEEkeywords}
Beamforming, minimum maximization, multicast communication, pinching-antenna systems.
\end{IEEEkeywords}

\section{Introduction}
The rapid advancement towards the sixth-generation (6G) wireless communication networks has been driven by demands for ultra-high data rates. 
Traditionally, network throughput is constrained by the fixed wireless channel, which is considered beyond human manipulation~\cite{1_rappaport2013millimeter}. 
However, in high-frequency bands such as millimeter-wave and terahertz, severe path loss and line-of-sight (LoS) blockage significantly degrade link reliability, especially over moderate to long transmission distances~\cite{4_sohrabi2017hybrid, 5_gong2020toward}.
To break this bottleneck and make the wireless environment more robust, recent research has focused on actively manipulating the wireless propagation environment through flexible-antenna techniques, such as reconfigurable intelligent surfaces (RISs) \cite{6_wu2021intelligent}, fluid-antennas \cite{7_wong2020fluid,Shojaeifard2022mimo,New2024tutorial}, and movable-antennas \cite{8_zhu2023movable,Zhu2025tutorial}.
Specifically, RIS modifies the wireless channel through programmable phase shifters~\cite{6_wu2021intelligent}. 
Moreover, fluid-antenna and movable-antenna systems dynamically adjust each antenna element's position within a spatial region to create favorable channel conditions~\cite{7_wong2020fluid,Shojaeifard2022mimo,New2024tutorial,8_zhu2023movable,Zhu2025tutorial}. 

Despite the recent progress in flexible-antenna systems, several inherent limitations restrict their practical applicability. First, the channel reconstruction is typically constrained within an aperture on the order of a few wavelengths, which limits their ability to mitigate large-scale path loss and establish robust LoS link, especially for cell-edge users. Second, once deployed, these systems generally
do not allow for flexible changes in antenna configuration, such as the number of antennas, which means that it is difficult to dynamically reconfigure the array size for different user demands or deployment conditions. 

To address these issues, the \emph{Pinching-Antenna SyStem (PASS)} has been proposed and experimentally demonstrated by NTT DOCOMO as a novel flexible-antenna technology~\cite{9_suzuki2022pinching}. 
By leveraging an arbitrarily long dielectric waveguide as a transmission medium, low-cost dielectric pinching antennas (PAs) can be attached or removed at any desired point along the waveguide. Unlike traditional flexible-antenna systems where antenna placement is confined to a limited aperture, the extended waveguide, whose length spans from a few meters to tens of meters, allows PA positions to be adjusted over a much larger spatial region. This dual flexibility in both antenna number and location allows PASS to establish strong LoS links for each user based on their individual channel conditions, which we refer to as \emph{``pinching beamforming''}. Moreover, the system is cost-effective and easy to deploy, as its operation relies only on mechanically adding or removing passive dielectric components. 

In essence, PASS can be viewed as a practical realization of the fluid-antenna and movable-antenna concepts proposed in prior works~\cite{New2024tutorial,Zhu2025tutorial}, while offering enhanced flexibility and scalability. In recognition of NTT DOCOMO's foundational contributions~\cite{9_suzuki2022pinching,11_yang2025pinching, 28_yuanwei2025architecture}, we refer to this technology as \emph{PASS} throughout this article. Moreover, PASS aligns with the emerging vision of \emph{surface-wave communication superhighways}~\cite{Wong2021vision}, which envisions leveraging in-waveguide propagation through reconfigurable waveguides to reduce path loss and improve signal power delivery~\cite{Liu2024,Chu2024}.

\subsection{Related Work} 
As a new paradigm of flexible-antenna technology, PASS has recently attracted considerable attention. Following the prototype demonstration in~\cite{9_suzuki2022pinching}, theoretical analyses have been conducted to explore the potential of PASS. 
Specifically, the authors in~\cite{10_ding2024flexible} analyzed the advantages of PASS over conventional fixed-location antenna systems in terms of ergodic sum rate.
The performance of PASS employing a single PA was studied in~\cite{31_Tyrovolas2025performance}, where closed-form expressions for outage probability and average rate were derived. 
The work in~\cite{12_ouyang2025array} focused on array gain characterization, where a closed-form upper bound was obtained and the existence of optimal antenna number and spacing was proven. 
For the PASS channel estimation problem, a sparse array-assisted multipath channel reconstruction method was proposed in~\cite{28_Guizhou2025channelesti} to estimate the PASS channel while reducing pilot overhead, and the authors in~\cite{27_Jianxiao2025channelesti} tackled the ill-conditioned and underdetermined characteristics of the PASS channel via a geometry-aware sparse recovery framework. 

Building on these theoretical exploration, PASS has been applied to various wireless communication settings. For the single-waveguide employment, uplink scenarios were exploited in~\cite{13_tegos2024minimum, 25_Tianwei2025uplink}, which focused on maximizing the minimum user rate through pinching beamforming design. 
Downlink transmission was investigated in~\cite{15_xu2024rate}, where the position of PAs was optimized to obtain the maximal channel gain. 
To further explore the potential of PASS in multiplexing, the multiple-waveguide architecture was considered for multiuser scenarios~\cite{26_zhaolin2025Modeling,25_bereyhi2025MIMOPASS, 17_guo2025deep, 24_Xiaoxia2025Ioint}.
Particularly, a physics-based power radiation model was proposed in~\cite{26_zhaolin2025Modeling}, based on which a joint beamforming optimization framework was developed to minimize transmit power.
A hybrid beamforming framework was proposed in~\cite{25_bereyhi2025MIMOPASS} to optimize joint beamforming via fractional programming. 
 The authors in~\cite{17_guo2025deep} developed a staged graph-based learning architecture to sequentially learn pinching beamformer and transmit beamformer. 
 In~\cite{24_Xiaoxia2025Ioint}, both optimization-based and learning-based methods were proposed for sum-rate maximization.

In addition to the aforementioned multiuser transmission scenarios, PASS-based multiple access technologies are introduced for interference managements.
The PA activation strategy under a non-orthogonal multiple access (NOMA) architecture was studied to enhance downlink throughput~\cite{14_wang2024antenna}, and a waveguide division multiple access (WDMA)-based multiple access model was introduced to enable spatial user access via dedicated waveguides~\cite{29_zhao2025WDMA}. 
Besides wireless communication scenarios, the application of PASS in integrated sensing and communication (ISAC) systems was studied in~\cite{30_zhangzheng2025ISAC}.

\subsection{Motivations and Contributions}
The aforementioned studies have validated the effectiveness of PASS in enhancing the performance of wireless communications. However, existing works have primarily focused on unicast transmission scenarios, where distinct private messages are delivered to individual users. In contrast, the application of PASS in multicast systems--where a common message is simultaneously transmitted to multiuser--remains largely unexplored.

Physical-layer multicast has emerged as a promising solution for efficient content delivery, particularly in the context of content-aware and group-based wireless communication paradigms. On the other hand, pinching beamforming offers the ability to establish stable and strong LoS links to each user by dynamically positioning antennas along the waveguide. This capability holds the potential to mitigate the classical performance bottleneck in multicast communications, which is often constrained by the user with the worst channel condition. 
These observations motivate the exploration of PASS for multicast communications, which is the focus of this article. The main contributions are summarized as follows:
\begin{itemize}
	\item We propose a PASS-enabled multicast framework where pinched dielectric waveguides are deployed to deliver a common message to multiuser within a service region. Based on this system model, we formulate a multicast beamforming problem that maximizes the multicast rate, i.e., the minimum achievable rate among all users, to ensure rate fairness and robust group performance.
	\item For the single-waveguide scenario, we derive a closed-form expression for the optimal PA location under the assumption of a single PA and linearly distributed users. Based on this result, we also derive a closed-form expression for the achievable multicast rate and prove that it strictly outperforms conventional fixed-location antenna systems. Furthermore, we derive a closed-form solution for the optimal PA placement under arbitrary user distributions by formulating and solving a Chebyshev center problem. For the multiple-PA case, we develop an element-wise alternating optimization (AO) algorithm to efficiently compute high-quality pinching beamformers with low computational complexity.
	\item For the case of multiple waveguides, we formulate a joint transmit and pinching beamforming optimization problem and address its non-convexity through a majorization-minimization (MM)-based AO framework. Within this framework, the pinching beamformer is updated using an element-wise sequential optimization method, where a surrogate objective function is constructed via MM approximation. Simultaneously, the transmit beamformer is designed by deriving a concave lower-bound approximation of the objective, which enables the use of second-order cone programming (SOCP) to efficiently obtain high-quality solutions.	
	\item We present numerical results to validate the performance advantage of PASS and the effectiveness of the proposed algorithms. The results demonstrate that: i) PASS achieves significantly higher multicast rates compared to conventional fixed-location antenna systems, including massive MIMO and hybrid analog-digital designs, especially as the number of users and spatial coverage increase; and ii) increasing the number of PAs further improves the multicast rate of PASS. These results confirm the scalability and potential of PASS for physical-layer multicast. 
\end{itemize}

\begin{figure*}[!t]
\centering
\includegraphics[height=0.25\textwidth]{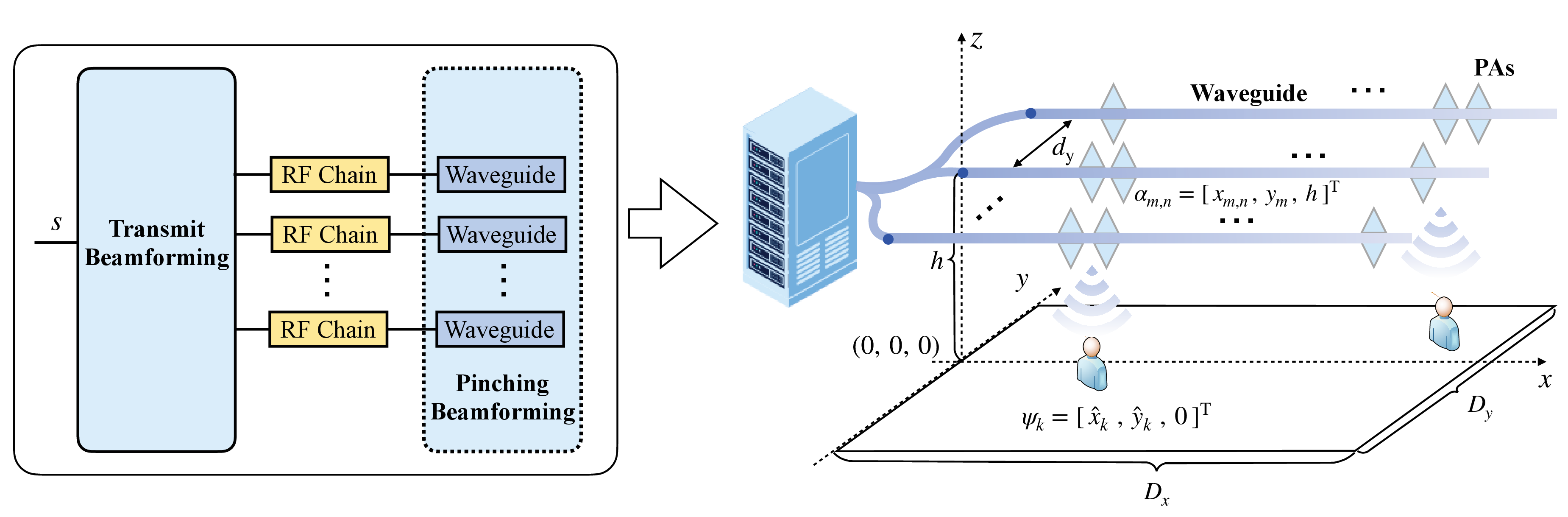}
\caption{Illustration of the PASS-enabled multicast communications.}\vspace{-2mm}
\label{Fig_1}
\vspace{-2pt}
\end{figure*}

\subsection{Organization and Notations}
The remainder of this paper is organized as follows. Section~\ref{System_Model} introduces the PASS-enabled system model and formulates the minimum-rate maximization problem. 
Section~\ref{single_waveguide} presents the pinching beamforming optimization for the single-waveguide deployment.
In Section~\ref{multiple_waveguide}, the joint beamformer is optimized for the multiple-waveguide setup. 
Numerical results are presented in Section~\ref{simulation}.
Finally, Section~\ref{conclusion} concludes this paper.

\emph{Notations:} 
Scalars, vectors, and matrices are represented by regular, bold lowercase, and  bold uppercase (e.g., $x$, $\mathbf {x}$, and $\mathbf{X}$) letters, respectively. The sets of complex and real numbers are denoted by $\mathbb{C}$ and $\mathbb{R}$, respectively. The inverse, conjugate, transpose, conjugate transpose, and trace operators are denoted by $(\cdot)^{-1}$, $(\cdot)^*$, $(\cdot)^{\rm T}$, $(\cdot)^{\rm H}$, and $\mathrm{Tr}(\cdot)$, respectively. ${\mathcal C}{\mathcal N}(a, b^2)$ is denoted as a circularly symmetric complex Gaussian distribution with mean $a$ and variance $b^2$, and ${\mathcal U}(a, b)$ is denoted as a uniform distribution over the interval $[a, b]$. The expectation operator is denoted by ${\mathbb{E}}\{\cdot\}$. The absolute value and Euclidean norm are denoted by $|\cdot|$ and $\|\cdot\|$, respectively. The real part of a complex number is denoted by $\Re \{\cdot\}$. The probability that the event $A$ occurs is denoted by $\Pr\{A\}$. The big-O notation is denoted by ${\mathcal O}(\cdot)$. 
\vspace{-5pt}
\section{System Model}\label{System_Model}
In this section, we propose a PASS-enabled multicast framework. As shown in {\figurename}~\ref{Fig_1}, the BS is equipped with $M$ waveguides, each equipped with $N$ PAs, to serve $K$ users. 
Let ${\mathcal K}$ denote the set of users. The location of each user $k\in {\mathcal K}$ is given by $\bm{\psi}_{k} = \left[\,{\hat x}_{k},\, {\hat y}_{k},\, 0\,\right]^{\rm T}$. 
The users are randomly distributed in a square region of size $D = D_{\mathrm{x}} \times D_{\mathrm{y}}$ $\text{m}^2$. 

We assume that all waveguides are deployed at a fixed height of $h$, and the PASS spans across the user region $D$. 
Let $\mathcal{M}$ and $\mathcal{N}_m$ denote the sets of all the waveguides and PAs deployed on the $m$th waveguide, respectively. The waveguides are uniformly spaced along the $y$-axis with an equal interval of $d_{\rm y} = D_{\rm y}/(M - 1)$ m.
Considering a waveguide fed by a dedicated radio frequency (RF) chain located at $\bm{\alpha}_{m,0} = \left[ \,0,\, y_m,\, h \, \right]^{\rm T}$, the position of the $n$th PA on the $m$th waveguide is denoted by $\bm{\alpha}_{m,n} = \left[\,x_{m,n},\, y_m,\, h\,\right]^{\rm T} $, $ n \in{\mathcal N}_m$ and $ m \in {\mathcal M}$. Here, $x_{m,n}$ represents the horizontal distance along the $x$-axis from the feed point to the $n$th PA, and $y_m$ is the $y$-coordinate of the $m$th waveguide. 
The PA locations along $x$-axis on the $m$th waveguide are collected into the vector $\mathbf{x}_m = [x_{m,1}, x_{m,2}, \dots, x_{m,N}]^{\rm T}\in {\mathbb R}^{N\times 1}$, where $0 \le x_{m,1} < x_{m,2} < \dots < x_{m,N} \le D_{\rm x}$. In addition, we fix the minimum inter-PA spacing as $ |x_{m,n} - x_{m,n-1}| \geq \Delta_{\rm min} = \lambda/2$, for $m\in {\mathcal M}$, $n\in {\mathcal N}_m$ and $n\neq 1$, to eliminate mutual coupling (MC) \cite{12_ouyang2025array}, where $\lambda$ is the free-space wavelength. The total PA locations along the $x$-axis can be stacked as ${\bf X} = \left[\,{\bf x}_1, \, {\bf x}_2, \cdots, {\bf x}_M\,\right]\in {\mathbb R}^{N \times M}$, which serves as the primary optimization variable in the pinching beamforming design. 
\subsection{Transmission Model}
Let $s\in {\mathbb C}$ denote the normalized information symbol for multicast communications with ${\mathbb E}[|s|^2] = 1$. The signal is first precoded by a transmit beamformer, which can be denoted as
	$\mathbf{w} \triangleq \bigl[\,\omega_{1},\, \omega_{2},\, \dots,\, \omega_{M}\,\bigr]^{\mathrm{T}}\in {\mathbb C}^{M\times 1}$. The precoded symbols are then upconverted by the RF chain and fed into each waveguide at $\bm{\alpha}_{m, 0}$.  Within the $m$th waveguide, each PA introduces a controllable in-waveguide phase shift to the input symbol, which can be expressed as follows:
\begin{align}
    {\bf g}({\bf x}_m) \triangleq\left[\rho_1{\rm e}^{-{\rm j}\theta(x_{m,1})}, \rho_2{\rm e}^{-{\rm j}\theta(x_{m,2})}, \cdots, \rho_n {\rm e}^{-{\rm j}\theta(x_{m,N})} \right]^{\rm T},
\end{align}
where $\rho_n$ denotes a power-scaling factor of the $n$th PA. For analytical simplicity, we assume equal power allocation across the $N$ PAs, which implies $\rho_n^2 = \rho = 1/N$, $ n \in \mathcal{N}$~\cite{26_zhaolin2025Modeling}.
The phase shift incurred from the feed point $\bm{\alpha}_{m, 0}$ to the $n$th PA is denoted by
\begin{align}
\theta(x_{m,n})
= 2\pi\,\frac{\|\boldsymbol{\alpha}_{m,0} - \boldsymbol{\alpha}_{m,n}\|}{\lambda_{g}} \triangleq \kappa_g x_{m,n},
\end{align}
where $\kappa_g = {2\pi}/{\lambda_g}$ denotes the guided wavenumber, and $\lambda_{g} = {\lambda}/{n_{\mathrm{eff}}}$ is the guided wavelength with $n_{\mathrm{eff}}$ representing the waveguide's effective refractive index~\cite{10_ding2024flexible}. 
Since each waveguide is connected to a subset of PAs, the in-waveguide channel matrix across the $M$ waveguides is a block-diagonal matrix, which can be written as follows:
\begin{align}
	\mathbf{G}\bigl(\mathbf{X}\bigr) & = \mathrm{blkdiag}\Bigl\{  {\bf g}\left(\mathbf{x}_{1}\right), \,{\bf g}\left(\mathbf{x}_{2}\right),\,\dots, {\bf g}\left(\mathbf{x}_{M}\right)\Bigr\}\notag \\
	& = \left[\begin{array}{cccc}
            {\bf g}\left({\bf x}_{1}\right) &  {\bf 0} & \dots &  {\bf 0}\\
             {\bf 0} &  {\bf g}\left({\bf x}_{2}\right) & \dots &  {\bf 0}\\
            \vdots & \vdots & \ddots & \vdots\\
             {\bf 0} &  {\bf 0} & \dots &  {\bf g}\left( {\bf x}_{M}\right)
            \end{array}\right].
\end{align}

We consider high-frequency bands where the free-space channel is well-approximated by the LoS-dominant model~\cite{10_ding2024flexible}.
Let ${\bf h}_{m,k}(\mathbf{x}_m)\in {\mathbb C}^{N\times 1}$ denote the channel vector between the $m$th waveguide and the user $k$. The overall PASS-enabled free-space channel vector can be written as follows:
\begin{align}
	{\bf h}_{k}({\bf X}) = \left[h_{1,k}(\mathbf{x}_1),\, h_{2,k}(\mathbf{x}_2),\, \cdots, \,h_{M,k}(\mathbf{x}_{M})\right]^{\rm T},
\end{align}
where each element is given by
\begin{align}
   \left[ h_{m,k}(\mathbf{x}_m) \right]_n
    = \frac{\sqrt{\eta}}{D_{k}(x_{m,n})}\,
    \exp\Bigl(-\,{\rm j} \kappa D_{k}(x_{m,n})\Bigr).
\end{align}
Here, $\eta=c^2/(16\pi^2f_c^2)$, $c$ is the speed of light, $f_c$ is the carrier frequency, and $\kappa = 2\pi/\lambda$ represents the free-space wavenumber. Additionally, $D_{k}(x_{m,n}) = \|\bm{\alpha}_{m,n} - \bm{\psi}_{k}\|$ denotes the distance between the $n$th PA of the $m$th waveguide and the user $k$. Consequently, the received signal at user $k$ can be expressed as follows:
\begin{align}\label{y_k}
    & y_{k} = {\bf h}_k^{\rm H}({\bf X}){\bf G}({\bf x})s + n_k,
\end{align}
where $n_{k}\sim\mathcal{CN}(0,\,\sigma^{2}_k)$ is the additive white Gaussian noise at user $k$ with the noise variance $\sigma^{2}_k$. 
The achievable rate of user $k$ can be written as follows:
\begin{align}
	R_{k}\bigl(\mathbf{X}, \mathbf{w}\bigr) = \log_{2}\Bigl(1 +\sigma_{k}^{-2} \bigl\lvert\mathbf{h}_k^{\mathrm{H}}\bigl(\mathbf{X}\bigr) \mathbf{G}\bigl(\mathbf{X}\bigr) \mathbf{w}\bigr\rvert^{2}\Bigr).
\end{align}
\vspace{-12pt}
\subsection{Problem Formulation}\label{Formulation}
Since all users receive the common multicast data symbol, the multicast rate is limited by the worst-case user, which is given as follows:
\begin{align}\label{obj_M>1}
	R_{\rm mc}\bigl({\bf X}, {\bf w}\bigr) = \mathop{\min}_{k\in {\mathcal K}}\;R_{k}\bigl({\bf X}, {\bf w}\bigr).
\end{align}
The maximization of the multicast rate can be formulated as follows:
\vspace{-5pt}\begin{subequations}\label{MW_obj_original}
\begin{align}
	{\mathcal P}_0: & \, \mathop{\max}_{\mathbf{X}, \mathbf{w}}\,R_{\rm mc}\bigl(\mathbf{X}, \mathbf{w}\bigr) \\
	        \text{s.t.} 
        & \quad 0 < x_{m,1} < x_{m,2} < \dots < x_{m,N} < D_{\rm x},\label{overlap} \\
        & \quad |x_{m,n} - x_{m,n-1}| \geq \Delta_{\rm min}, \forall m\in {\mathcal M},\, n\neq 1, \label{CM} \\
        & \quad {\rm Tr}\left({\bf w}{\bf w}^{\rm H}\right) \leq P_{\rm t},\label{power_constraint} \\
        & \quad {\bf X}\in {\mathcal X}\label{candidate_set},
\end{align}
\end{subequations}
where the constraints in~\eqref{overlap} indicate that each PA must remain along the waveguide in a strictly increasing order to avoid overlaping, $\Delta_{\rm min}$ in \eqref{CM} represents the minimum inter-PA spacing to eliminate MC effects, the transmit power constraint is specified in~\eqref{power_constraint}, and $\mathcal{X}$ represents the candidate set of the PA locations along the $x$-axis.

\section{Single-Waveguide Scenario}\label{single_waveguide}

\begin{figure}[!t]
\centering
\includegraphics[height=0.2\textwidth]{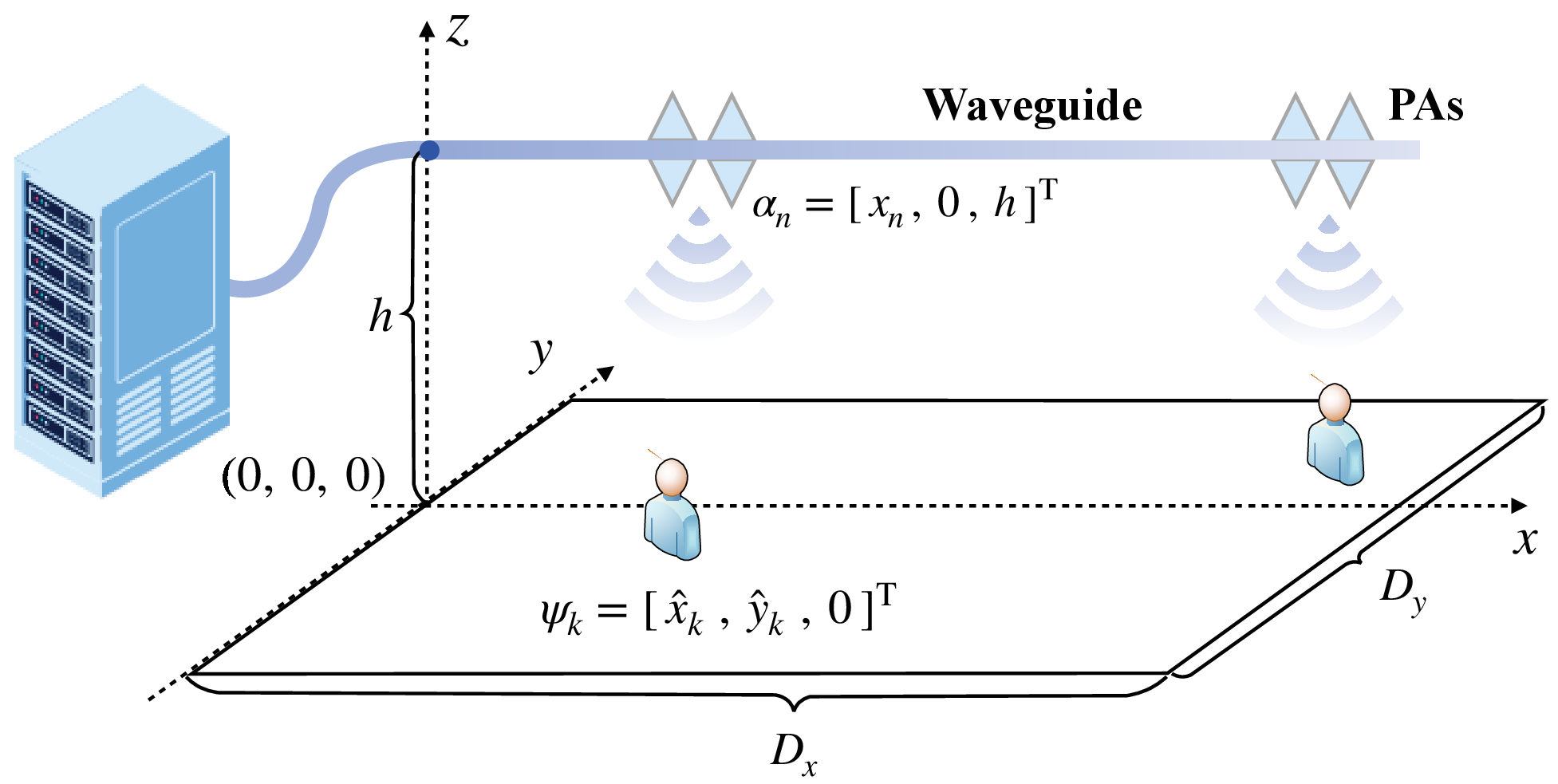}
\caption{Illustration of the single-waveguide based PASS.}\vspace{-5mm}
\label{Fig1_Schematic}
\end{figure}

To gain fundamental insights into the system design, we begin by analyzing the single-waveguide case with only one pinched waveguide. As shown in {\figurename}~\ref{Fig1_Schematic}, the single waveguide is deployed parallel to the $x$-axis at the fixed height $h$.
In this case, the location of the PA can be represented as ${\bm\alpha}_n = \left[\,x_n, \,0, \,h\,\right]^{\rm T}$, and the pinching beamformer can be simplified as ${\bf g}({\bf x})$ with ${\bf x} = \left[x_1, x_2, \cdots, x_N\right]^{\rm T}$.  
We further denote the distance between ${\bm \psi}_k$ and ${\bm \alpha}_n$ as $D_{n,k}(x_n) = \|\bm{\alpha}_{n} - \bm{\psi}_{k}\|$. Finally, the received signal at user $k$ can be expressed as follows:
\vspace{-5pt}\begin{align}
    & y_{k} = {\bf h}_k^{\rm H}({\bf x}){\bf g}({\bf x})s + n_k \nonumber\\
    &= \sum_{n=1}^{N} \frac{\sqrt{\eta \rho}\exp(-{\rm j}(\kappa D_{k}(x_n)+\kappa_g x_n))}{D_{k}(x_n)}s +n_{k}.
\end{align}
Consequently, the signal-to-noise ratio (SNR) at user $k$ is given by
\begin{align}
    \mathrm{SNR}_{k}(\mathbf{x})
    &= \frac{\eta P_{\rm t}\left\lvert \displaystyle\sum\limits_{n=1}^{N} 
    \frac{\exp(-{\rm j}(\kappa D_{k}(x_n)+\kappa_g x_n))}{D_{k}(x_n)}\right\rvert^{2}}
    {N  \sigma^{2}_k}.
\end{align}
Then, the achievable rate of user $k$ can be written as follows:
\begin{align}
    R_{k}(\mathbf{x})
    = \log_{2}\bigl(1 + \mathrm{SNR}_{k}(\mathbf{x})\bigr),
\end{align}
and the overall multicast rate can be expressed as follows: 
\begin{align}
    R_{\mathrm{mc}}(\mathbf{x}) 
    = \mathop\min_{k \in {\mathcal K}} \; R_{k}(\mathbf{x}).
\end{align}

Accordingly, the problem defined in~\eqref{MW_obj_original} simplifies to the following:
\begin{subequations}
    \begin{align}
       \quad{\mathcal P}_1: & \, \, \, \underset{\mathbf{x}}{\rm{max}} \quad R_{\mathrm{mc}}(\mathbf{x})\label{obj_function}\\
        \text{s.t.} 
        & \quad 0 \le x_{1} < x_{2} < \dots < x_{N} < D_{\rm x},\label{x_overlap} \\
        & \quad |x_{n} - x_{n-1}|\geq \Delta_{\rm min},\, n\neq 1, \label{mutual_coupling} \\
        & \quad {\bf X}\in {\mathcal X}.\label{candi_P1}
    \end{align}
    \label{eq:maxminObjective}
\end{subequations}\vspace{-8mm}

\subsection{Single-PA Case}
We first consider the scenario that a single PA is activated along the waveguide. We denote the PA position as $x$. Therefore, the SNR of user $k$ can be simplified as follows:
\begin{align}
    \mathrm{SNR}_{k}(x)
    &= \frac{\eta P_{\rm t}\left\lvert
    {{\rm exp}\left(-{\rm j}\left(\kappa D_k(x)+\kappa_g x\right)\right)}
    \right\rvert^{2}}
    {D_k^2(x)\sigma^{2}_k},
\end{align}
where the distance between the PA and user $k$ is given by
\begin{align}\label{freespace_dis}
    D_k(x) = \sqrt{h^2+(x-{\hat x}_k)^2+{\hat y}_k^2}.
\end{align}
Defining $C_k \triangleq P_{\rm t}\eta/\sigma_k^2$, we rewrite the multicast rate as follows:
\begin{align}
    R_{\rm mc}(x) = \underset{k\in {\mathcal K}}{\rm{min}} \, R_k(x),
\end{align}
where
\begin{align}
	R_k(x) = \log_2\left(1+\frac{C_k}{D_k^2(x)}\right).
\end{align}

To determine the optimal PA location along the $x$-axis, we adopt a candidate-point search approach. We begin by establishing the unimodality of $R_{\rm mc}(x)$. Particularly, the per-user rate function $R_k(x)$ is strictly unimodal, as can be seen from its derivative:
\begin{align}
    \frac{dR_k(x)}{dx} = \frac{-2C_k(x-\hat{x}_k)}{\ln 2\,\Bigl(D_k^2(x)+C_k\Bigr)\,D_k^2(x)}.
\end{align}
It is evident that there exists a unique maximum of $R_k(x)$ at $x = \hat{x}_k$. 
 Furthermore, as the lower envelope of $\{R_k(x)\}_{k=1}^{K}$,  $R_{\rm mc}(x)$ is also strictly unimodal~\cite{Unimodality}.
As a result, the global maximizer of $R_{\rm mc}(x)$ must lie at a point where either:
\begin{itemize}
    \item $x = \hat{x}_k$, $k\in {\mathcal K}$, which correspond to the maximum of a per-user rate $R_k(x)$; or
    \item $x = x^{\rm int}_{ij}$, where $x^{\rm int}_{ij}$ satisfies $R_i(x^{\rm int}_{ij}) = R_j(x^{\rm int}_{ij})$ for some $i \neq j$.
\end{itemize}

Accordingly, the candidate set can be constructed as follows:
\begin{align}
\mathcal{X}_{\rm sp} = \mathcal{X}_1 \cup \mathcal{X}_2, 
\end{align}
where $\mathcal{X}_1\triangleq\{\hat{x}_1, \dots, \hat{x}_K\}$ and
\begin{align}
	\mathcal{X}_2\triangleq \{x^{\rm int}_{ij} \,|\, R_i(x) = R_j(x),\; 1\leq i<j\leq K\}.
\end{align}
Finally, we evaluate $R_{\rm mc}(x)$ at all candidate points in $\mathcal{X}_{\rm sp}$, and the optimal PA location can be selected  as follows:
\begin{align}\label{cp_obj}
 x^{\rm opt} = \underset{x \in \mathcal{X}_{\rm sp}}{\arg\max} \, R_{\rm mc}(x).
\end{align}

{\bf Complexity Analysis}: Since the cardinality of the set $\mathcal{X}_{\rm sp}$ is given by $1+K(K-1)/2$, the complexity of the proposed method scales as $\mathcal{O}(K^2)$. In contrast, the complexity of exhaustive search scales as $\mathcal{O}({L})$, where $L$ denotes the discrete candidate points in the one-dimensional search. Therefore, for typical system settings where $K$ is moderate but the search resolution is fine-grained, the proposed method achieves the optimal performance with significantly lower complexity.

We next consider a special case to reveal more design insights, where the users are linearly distributed parallel to the $x$-axis. Additionally, all users are assumed to share identical system parameters, including the noise power $\sigma^2$, channel gain factor $\eta$, and a common $y$-axis coordinate $\hat{y}$. Under this assumption, we have $C_k = C$ for all users. The per-user rate then simplifies to
\begin{align}
   & R_k(x) = \log_2\left(1+\frac{C}{D_k^2(x)}\right),
   \end{align}
where
\vspace{-0.5em}
\begin{align}
D_k^2(x) = h^2 + (x - \hat{x}_k)^2 + \hat{y}^2.
\end{align}
Accordingly, the multicast rate can be rewritten as follows:
\begin{align}
    R_{\rm mc}(x) = \log_2\left(1+\frac{C}{\max_{k} D_k^2(x)}\right).
\end{align}
In this case, minimizing $\max_k D_k^2(x)$ is equivalent to minimizing the maximum squared distance between the PA and all users, which corresponds to solving a Chebyshev center problem along the $x$-axis~\cite{Chebyshev}.

The function $\max_k D_k^2(x)$ forms the upper envelope of $K$ convex parabolic functions. 
Its minimum is achieved at the position where the distances from the PA to the leftmost and rightmost users are identical~\cite{Boyd}.
Specifically, by solving
\begin{align}
    (x-\hat{x}_{\min})^2 = (x-\hat{x}_{\max})^2,
\end{align}
the optimal PA location can be obtained as follows:
\begin{align}
    x^{\rm opt} = \frac{\hat{x}_{\min} + \hat{x}_{\max}}{2}.
\end{align}
where $\hat{x}_{\min}$ and $\hat{x}_{\max}$ represent the $x$-coordinate of the leftmost and rightmost users, respectively. 
Therefore, the global optimum is achieved by positioning the PA at the midpoint between the two boundary users, which minimizes the maximum distance, i.e., achieves the maximum multicast rate.

By substituting $x^{\mathrm{opt}}$ into $R_{k}(x)$ and denoting $\Delta_x\triangleq \hat{x}_{\rm max} - \hat{x}_{\rm min}$ as the distance between the leftmost and rightmost users, we obtain a closed-form upper bound on the achievable multicast rate as follows:
\begin{align}\label{R_min_special}
	R_{\mathrm{mc}}({x}) 
	&= 
	\log_2\left( 1 
	+ \frac{\eta P_{\rm t}}{ \sigma^2\left(\frac{\Delta_x^2}{4}+h^2+{\hat y}^2\right)} \right).
\end{align}
\begin{lemma}\label{lemma1}
Under the linear user distribution, the average multicast rate achieved by a single PA can be calculated as follows:
\begin{align}\label{C_N1M12K}
	C_{\rm PASS} & = \frac{2}{D_{\rm x}^2 \ln 2} \Bigl( J(A + P_{\rm eff}) - J(A) \Bigr),
\end{align}
where
\begin{align}
	& J(C) = D_{\rm x}\, I_1(C) - I_2(C)
	\end{align}
	with
	\begin{align}
&  I_1(C) = 2\left( T \ln(T^2 + C) - 2T + 2\sqrt{C}\, \arctan\left( \frac{T}{\sqrt{C}} \right) \right),\notag \\
& I_2(C) = 2\left( ({T^2 + C}) \ln(T^2 + C)
- {T^2}
- {C} \ln C \right),
\end{align}
and
\begin{align}
	A = h^2 + \hat{y}^2, \quad
P_{\rm eff} = \frac{\eta P_{\rm t}}{\sigma^2}, \quad
T = \frac{D_{\rm x}}{2}.
\end{align}
\end{lemma}
\begin{IEEEproof}
Please refer to Appendix~\ref{proof-lemma1} for more details.
\end{IEEEproof}

 To gain further insights from the simplified model, we compare the performance of the PASS with that of a conventional fixed-location antenna system in the high-SNR regime. Under this assumption, the average multicast rate of the PASS in~\eqref{C_N1M12K} can be approximated as follows:
  \begin{align}\label{PASS_appro}
 	C_{\rm PASS} \approx {\rm log}_2\!\Bigl(\frac{P_{\rm eff}}{A}\Bigr) \;-\; \frac{D_{\rm x}^2}{24\,A\,\ln2}.
\end{align}

We next consider a fixed-location antenna system, where a single antenna is positioned at
$[\, D_{\rm x}/2, \, 0, \, h\,]^{\rm T}$. In this case, the average multicast rate in the high-SNR regime can be given as follows:
\begin{align}\label{Conv_appro}
	C_{\rm Conv} \approx {\rm log}_2\!\Bigl(\frac{P_{\rm eff}}{A}\Bigr) \;-\; {\frac{KD_{\rm x}^2}{4A\,(K+2)\ln2}}.
\end{align}
The full derivation of \eqref{PASS_appro} and \eqref{Conv_appro} is provided in Appendix~\ref{proof-Theorem1}.

The corresponding multicast rate gain of the PASS over the fixed-location antenna system is measured through the rate difference:
\begin{align}\label{delta_R}
	\Delta C = C_{\rm PASS} - C_{\rm Conv} = \frac{D_{\rm x}^2}{A\,\ln2}\left(\frac1{24} - \frac{K}{4(K+2)}\right).
\end{align}
This leads to the following conclusion.
\begin{remark}\label{Theorem1}
In the high-SNR regime, the average multicast rate achieved by the PASS employing a single PA is strictly higher than that of a conventional fixed-location antenna system. Moreover, the corresponding multicast rate gain increases monotonically with $D_{\rm x}^2$.
 \end{remark}

\subsection{Multiple-PA Case}\label{M=1N>1}
We now consider the configuration where multiple PAs are deployed over the waveguide, which gives rise to the pinching beamforming optimization problem in \eqref{eq:maxminObjective}. 
Obtaining the globally optimal solution would require an exhaustive search over all feasible PA placements, which is computationally intractable.

To overcome this challenge, we propose an element-wise AO approach to solve problem~\eqref{eq:maxminObjective}, where each $x_{n}$ is optimized sequentially by fixed the others.
Specifically, the subproblem for optimizing each $x_n$ can be equivalently rewritten as follows:
\begin{align}\label{SNR_obj}
{\mathcal P}_2:\underset{{x}_n \in \mathcal{X}_{\rm mp}}{\max} \mathop\min_{k\in {\mathcal K}} \,\mathrm{SNR}_k({x}_n), \quad \text{s.t.} \eqref{mutual_coupling}.
\end{align}
Since the problem reduces to optimizing a single variable over a bounded interval, it can be efficiently addressed via a low-complexity one-dimensional grid search. Specifically, PAs are constrained to be placed at discrete preconfigured locations, and the optimal $x_n$ is obtained by evaluating the objective across all candidate points.

Using an $L$-point uniform grid over the interval $[\,0,\, D_{\rm x}\,]$, the feasible set is defined as follows:
\begin{align}
	\mathcal{X}_{\rm mp} = \left\{0, \frac{D_{\rm x}}{L-1}, \frac{2D_{\rm x}}{L-1}, \dots, D_{\rm x}\right\},
\end{align}
Furthermore, the partial objective function of~\eqref{SNR_obj} can be expressed as follows:
\begin{align}\label{obj_original}
\quad &\underset{{x}_n \in \mathcal{X}_{\rm mp}}{\max} \ \mathop\min\limits_{k \in {\mathcal K}}\left\{\frac{\eta P_{\rm t}\left|S_k^{n-}+A_{n,k}(x_n)\right|^2}{\sigma^{2}_k}\right\},
\end{align}
where
\begin{align}
S_k^{n-} &= \sum_{q \neq n}^{N} \frac{\exp\Bigl(-\,{\rm j}\,\left(\kappa\,D_{k}(x_q)+\kappa_g x_q\right)\Bigr)}{D_{k}(x_q)}
\end{align}
represents the aggregate contribution of the fixed PAs, and
\begin{align}
A_{k}(x_n) &= \frac{\exp\Bigl(-\,{\rm j}\,\left(\kappa\,D_{k}(x_n)+\kappa_g x_n\right)\Bigr)}{D_{k}(x_n)}
\end{align}
represents the contribution from the $n$th PA. It follows that
\begin{align}
& \bigl|S_k^{n-} + A_k(x_n)\bigr|^2\notag \\
\quad= &|S_k^{n-}|^2 + \frac{1}{D_{k}^2(x_n)} + 2\,\Re\!\left\{S_k^{n- *}\,\frac{e^{-{\rm j}\phi_k(x_n)}}{D_{k}(x_n)}\right\},
\end{align}
where $\phi_k(x_n) = \kappa D_{k}(x_n) + \kappa_g x_n$.  
Thus, the optimization for $x_n$ simplifies to the following:
\begin{align}\label{obj_final}
 & x_n^{\rm opt} = \underset{x_n \in \mathcal{X}_{\rm mp}}{\arg\max} \  {\widetilde R}_{\rm mc}(x_n), 
\end{align}
where ${\widetilde R}_{\rm mc}(x_n) = \min_{k}{\widetilde R}_{k}(x_n)$, and 
\begin{align}
	{\widetilde R}_{k}(x_n) =\left( \frac{1}{D_{k}^2(x_n)} + 2\,\Re\frac{\left\{S_k^{n- *} e^{-{\rm j}\phi_k(x_n)}\right\}}{D_{k}(x_n)} \right).
\end{align}
The optimization proceeds by sequentially updating each $x_n$ across all $N$ PAs in an iterative manner until convergence. The detailed element-wise AO-based algorithm for solving problem \eqref{obj_final} is given in Algorithm \ref{alg:greedy}.
\begin{algorithm}[t]
\caption{Element-wise AO-based Algorithm for Solving~\eqref{eq:maxminObjective}}
\label{alg:greedy}

\begin{algorithmic}[1]
\STATE initialize the optimization variables
\REPEAT
    \FOR{$\{n = 1,2,\dots,N\}$}
    \STATE {update} $x_n$ by solving problem~\eqref{obj_final} through one-dimensional search
    \ENDFOR
\UNTIL {the increase of the objective value of problem~\eqref{obj_function} falls below a predefined threshold}
\end{algorithmic}
\end{algorithm}

\subsection{Complexity and Convergence Analysis}
\subsubsection{\bf Complexity Analysis}
Let $L$ denote the number of discrete candidate locations. An exhaustive search would require a total complexity of ${\mathcal O}(KL^N)$. In contrast, the proposed element-wise AO-based approach has a complexity of ${\mathcal O}(I_{\rm EW}KN)$ with $I_{\rm EW}$ denoting the number of element-wise AO iterations.

\subsubsection{\bf Convergence Analysis}
At each iteration, the element-wise AO-based approach updates $x_n$ generates a monotonic sequence $\left\{{\widetilde R}_{\rm mc}\left({x}_n^{(t+1)}\right)\right\}$.
Since $\mathcal{X}_{\rm mp}$ is a finite discrete set and $R_{\rm mc}(\mathbf{x})$ is upper bounded by $\log_2\left(1 + P_{\rm t} \sum_n \eta/(\sigma^2 h^2)\right)$, the sequence of objective values is monotonically non-decreasing and must converge in a finite number of iterations. 

\section{Multiple-Waveguide Scenario}\label{multiple_waveguide}
We now design the joint transmit and pinching beamforming in the multiple-waveguide scenario. 

\subsection{MM-Based AO Framework}
The max-min problem in~\eqref{MW_obj_original} is non-convex and challenging to solve directly. To address this, we adopt the MM technique to transform the original objective function into a more tractable surrogate problem, in which the AO framework can be developed to update the transmit and pinching beamformer iteratively.

To facilitate analysis, we define the effective channel gain of each user $k\in {\mathcal K}$ as follows:
\begin{align}
	{g}_{{\rm eff}, k}({\bf X}, {\bf w})\triangleq{\bf h}_k^{\rm H}({\bf X}){\bf G}({\bf X}){\bf w}.
\end{align}
Based on the above expression, we construct a MM surrogate for the per-user rate as follows:
\begin{lemma}\label{lemma_surrogate}
Let $\left\{{\bf X}^{(t)},{\bf w}^{(t)}\right\}$ denote the solutions obtained at the $t$th iteration. Then $R_{k}\left({\bf X}, {\bf w}\right)$ is minorized by the following concave surrogate: 
\begin{align}
  & \widetilde{R}_{k}\left({\bf X}, {\bf w}|{\bf X}^{(t)}, {\bf w}^{(t)}\right) \notag \\
  & = {{ c}_k} + 2\Re \left\{ a_k{g}_{{\rm eff}, k}({\bf X}, {\bf w})\right\} - b_k{g}^2_{{\rm eff}, k}({\bf X}, {\bf w})\notag \\
 & \leq R_{k}\left({\bf X}, {\bf w}\right),\label{eq:Rate-surrogate}
\end{align}
where the coefficients are given by 
\begin{subequations}
\begin{align}
	& a_k = \sigma_k^{-2}{g}^*_{{\rm eff}, k}\left({\bf X}^{(t)}, {\bf w}^{(t)}\right),  \\
	& b_k = \sigma_k^{-2}{{g}^2_{{\rm eff}, k}\left({\bf X}^{(t)}, {\bf w}^{(t)}\right)\left(\sigma_k^2+{g}^2_{{\rm eff}, k}\left({\bf X}^{(t)}, {\bf w}^{(t)}\right)\right)^{-1}},  \\
    & {{ c}_k} = R_k\left({\bf X}^{(t)}, {\bf w}^{(t)}\right)-2b_k\sigma_k^2  -b_k{g}^2_{{\rm eff}, k}\left({\bf X}^{(t)}, {\bf w}^{(t)}\right).
\end{align}
\end{subequations}
 \end{lemma}

\textbf{\textit{Proof: }}Please refer to Appendix \ref{proof-lemma2} for more details.

The surrogate function $\widetilde{R}_{k}\left(\mathbf{X},\mathbf{w}|\mathbf{X}^{(t)},\mathbf{w}^{(t)}\right)$ is biconcave of $\mathbf{X}$ and $\mathbf{w}$,
as it is concave in ${\bf X}$ for fixed ${\bf w}$, and concave in ${\bf w}$ for fixed ${\bf X}$~\cite{21_gorski2007biconvex}.
This biconcave structure facilitates the use of the AO framework to alternately update $\mathbf{X}$ and $\mathbf{w}$ in a decoupled manner.

Based on the above analysis, the objective in~\eqref{obj_M>1} can be transformed into the following surrogate function: 
\begin{align}\label{eq:Problem-MM}
& {\widetilde R}_{\rm mc}\bigl({\bf X}, {\bf w}\bigr) = \mathop{\min}_{k\in {\mathcal K}}\;{\widetilde R}_{k}\left({\bf X}, {\bf w} |{\bf X}^{(t)}, {\bf w}^{(t)} \right).
\end{align}

\subsection{Pinching Beamforming Design}

In this subsection, we focus on optimizing the pinching beamformer ${\bf G}\left({\bf X}\right)$ with a given ${\bf w}$. The surrogate rate expression $\widetilde{R}_{k}\left({\bf X}|{\bf X}^{(t)}\right)$ can be rewritten as follows:
\begin{align}
& \widetilde{R}_{k}\left({\bf X}|{\bf X}^{(t)}\right) =c_{k} + 2\Re\left\{{\bf a}_{k}^{\rm H}\left({\bf X}\right){\bf w}\right\} - {\bf w}^{\rm H}{\bf H}_{{\rm eff},k}\left({\bf X}\right){\bf w},\label{surrogate}
\end{align}
where ${\bf a}_{k}=a_{k}{\bf h}_{{\rm eff}, k}^{\rm H}\left({\bf X}\right)$ with ${\bf h}_{{\rm eff}, k}\left({\bf X}\right) \triangleq {\bf G}^{\rm H}\left({\bf X}\right){\bf h}_{k}\left({\bf X}\right)$,
and 
\begin{align}
	& {\bf H}_{{\rm eff},k}\left({\bf X}\right) = b_k{\bf h}_{{\rm eff}, k}\left({\bf X}\right){\bf h}_{{\rm eff}, k}^{\rm H}\left({\bf X}\right).
\end{align}
Similar to the single-waveguide scenario, we adopt the element-wise sequential optimization to solve each $x_{m,n}$.
Upon~\eqref{eq:Problem-MM} and~\eqref{surrogate}, the subproblem for the optimization of $x_{m,n}$ can be formulated as follows:
\begin{align}\label{MM_X}
& {\mathcal P}_3: \mathop{\max}\limits _{x_{m,n}\in {\mathcal X}_{\rm mp}} {\widetilde R}_{\rm mc}(x_{m,n}), \quad
	        \text{s.t.}\, \eqref{CM},
\end{align}
where
\begin{align}\label{term1}
	& {\widetilde R}_{\rm mc}(x_{m,n}) \notag \\
	& = \mathop\min_{k\in {\mathcal K}}\left\{2\Re\left\{ \mathbf{a}_{k}^{\mathrm{H}}\bigl(x_{m,m}\bigr)\mathbf{w}\right\} - \mathbf{w}^{\mathrm{H}}\mathbf{H}_{{\rm eff},k}\bigl(x_{m,m}\bigr)\mathbf{w}\right\}.
\end{align}
To further reduce computational complexity of each iteration, we derive a partial MM-based surrogate objective function by isolating the dependency on $x_{m,n}$. 
Specifically, the term $\mathbf{a}_{k}^{\mathrm{H}}(x_{m,n})\mathbf{w}$ in~\eqref{term1} can be expanded as follows:
\begin{align}
	& \mathbf{a}_{k}^{\mathrm{H}}(x_{m,n})\mathbf{w} = a_k\sum_{m = 1}^{M}\left(w_m\sum_{n = 1}^{N}\Bigl
	(A_{m,n,k}(x_{m,n}) + S_k^{m,n-}\Bigr)\right)\notag \\
	& = a_kw_m A_{m,n,k}(x_{m,n}) + a_k\sum_{m = 1}^{M}\sum_{q \neq n}^{N}w_mS_k^{m,n-},
\end{align}
where
\begin{align}
	S_k^{m,n-} = \sum\limits_{p,q\neq m,n}^{NM}\frac{\rho \,{\rm exp}\left(-{\rm j}\bigl(\kappa D_k(x_{p,q})+\kappa_g\, x_{p,q}\bigr)\right)}{D_k(x_{p,q})}
\end{align}
is the aggregate contribution of the fixed PAs and
\begin{align}
	A_{m,n,k}(x_{m,n}) = \frac{\rho\, {\rm exp}\left(-{\rm j}\Bigl(\kappa D_k(x_{m,n})+\kappa_g\, x_{m,n}\Bigr)\right)}{D_k(x_{m,n})}
\end{align}
denotes the contribution of the $n$th PA employed on the $m$th waveguide.

Meanwhile, the term $\mathbf{w}^{\mathrm{H}}\mathbf{H}_{{\rm eff},k}(x_{m,n})\mathbf{w}$ in \eqref{term1} can be expressed as follows:

\begin{align}
 	& \mathbf{w}^{\mathrm{H}}\mathbf{H}_{{\rm eff},k}(x_{m,n})\mathbf{w}\notag \\
 	 & = b_k\left(\sum_{m = 1}^{M}\left(w_m\sum_{n = 1}^{N}\left(A_{m,n,k}(x_{m,n}) + S_k^{m,n-}\right)\right)\right)^2\notag  \\
 	 & = a_{m,n,k}A_{m,n,k}^2(x_{m,n}) + b_{m,n,k}A_{m,n,k}(x_{m,n}) + {c}_{m,n,k},
\end{align}

where
\begin{subequations}
\begin{align}
& a_{m,n,k} = b_kw_m^2,  \\
& b_{m,n,k} =  2b_kw_m \sum_{m = 1}^{M}\sum_{q \neq n}^{N}w_mS_k^{m,n-}, \\
& {c}_{m,n,k} = b_k\left(w_m\sum_{m = 1}^{M}\sum_{q \neq n}^{N}w_mS_k^{m,n-}\right)^2.
\end{align}
\end{subequations}

Taken together, the optimization of $x_{m,n}$ in the element-wise sequential optimization framework is formulated as \eqref{MM_greedy_search}, which is shown on the top of the next page. Then, a one-dimensional search can be performed to update $x_{m,n}$. 

\begin{figure*}[!t]
\begin{align}\label{MM_greedy_search}
{\mathcal P}_4: \mathop{\max}\limits _{x_{m,n}\in {\mathcal X}_{\rm mp}} & \mathop\min_{k\in {\mathcal K}} \Bigg\{
   a_{m,n,k}A_{m,n,k}^2\bigl(x_{m,n}\bigr)
   \;+\;
   2 \Re \Bigl\{a_k w_m A_{m,n,k}\bigl(x_{m,n}\bigr)\Bigr\}
   \;+\;
   b_{m,n,k}A_{m,n,k}\bigl(x_{m,n}\bigr)\Bigg\},\quad \text{s.t.} \quad \eqref{CM}.
\end{align}
\hrulefill
\end{figure*}

\begin{algorithm}[t]
\caption{MM-based AO Algorithm for Joint Beamforming Problem~\eqref{obj_M>1}}
\label{algo:AO}
\begin{algorithmic}[1]
\STATE initialize the optimization variables, define $\epsilon$ as a predefined threshold
\REPEAT
    \STATE  update $\mathbf{X}^{(t+1)}$ by solving \eqref{MM_greedy_search} 
    \STATE update $\mathbf{w}$ by solving \eqref{eq:Problem-F-socp}
\UNTIL $\bigl\|\mathbf{X}^{(t+1)} - \mathbf{X}^{(t)}\bigr\| \le \epsilon$ and $\bigl\|\mathbf{w}^{(t+1)} - \mathbf{w}^{(t)}\bigr\| \le \epsilon$
\end{algorithmic}
\end{algorithm}

\subsection{Transmit Beamforming Design}
In this subsection, we aim to optimize the transmit beamforming vector ${\bf w}$
with a given pinching beamformer ${\bf G}\left({\bf X}\right)$. From~\eqref{eq:Rate-surrogate}, the surrogate rate function $\widetilde{R}_{k}\left(\mathbf{X},\mathbf{w}|\mathbf{X}^{(t)},\mathbf{w}^{(t)}\right)$ can be expressed as a quadratic function of $\mathbf{w}$: 
\begin{align}
\widetilde{R}_{k}\left(\mathbf{w}|\mathbf{w}^{(t)}\right) & =c_{k}+2\Re\left\{ a_{k}\mathbf{h}_{{\rm eff}, k}^{\mathrm{H}}\mathbf { w}\right\} -b_{k}|\mathbf{h}_{{\rm eff}, k}^{\mathrm{H}}\mathbf { w}|^{2}\nonumber \\
\label{eq:Rate-surrogate-F}
\end{align}

By substituting~\eqref{eq:Rate-surrogate-F} into the MM surrogate formulation~\eqref{eq:Problem-MM}, the transmit beamforming subproblem becomes
\begin{subequations}
\begin{align}
& {\mathcal P}_5: \mathop{\max}\limits _{\mathbf{w}}  \mathop\min_{k \in {\mathcal K}}\left\{2\Re\left\{ a_{k}\mathbf{h}_{{\rm eff}, k}^{\mathrm{H}}\mathbf { w}\right\} -b_{k}|\mathbf{h}_{{\rm eff}, k}^{\mathrm{H}}\mathbf { w}|^{2}\right\}  \\
& {\rm s.t.} \quad {\rm Tr}\left({\bf w}{\bf w}^{\rm H}\right) \leq P_{\rm t}.\label{eq:Problem-F-1}
\end{align}
\end{subequations}
To address the pointwise minimum in the objective, we introduce an auxiliary variable ${\gamma}$ and reformulate the problem as follows: 
\begin{subequations}\label{eq:Problem-F-socp}
\begin{align}
&{\mathcal P}_6:\mathop{\max}\limits _{\mathbf{w},{\gamma}}  \;\;\gamma \\
&{\rm s.t.} \ {\rm Tr}\left({\bf w}{\bf w}^{\rm H}\right) \leq P_{\rm t}, \\
 &\quad\,\,\, 2\Re\left\{ a_{k}\mathbf{h}_{{\rm eff}, k}^{\mathrm{H}}\mathbf { w}\right\} -b_{k}|\mathbf{h}_{{\rm eff}, k}^{\mathrm{H}}\mathbf { w}|^{2}\geq\gamma, \\
 & \quad\,\,\,\,\forall k\in\mathcal{K}.
\end{align}
\end{subequations}
Problem (\ref{eq:Problem-F-socp}) is an SOCP problem and can be efficiently solved using convex solvers~\cite{22_CVX2018}, such as MOSEK \cite{23_Mosek2018}.

\subsection{Complexity and Convergence Analysis}
The proposed MM-based AO algorithm for solving~\eqref{MW_obj_original} is summarized in Algorithm~\ref{algo:AO}.
\subsubsection{\bf Complexity Analysis}
The element-wise sequential optimization for updating ${\bf X}$ has a complexity of ${\mathcal O}(I_{\rm EW}KMN)$. The complexity of SOCP for solving problem~\eqref{eq:Problem-F-socp} is $\mathcal{O}(M+M^{3}+MK^{3.5}+M^{3}K^{2.5})$. Therefore, the computational complexity of Algorithm~\ref{algo:AO} scales as ${\mathcal O}(I_{\rm AO}(I_{\rm EW}KMN + M+M^{3}+MK^{3.5}+M^{3}K^{2.5}))$.

\subsubsection{\bf Convergence Analysis}
The surrogate function~\eqref{eq:Problem-MM} generates a monotonically non-decreasing sequence of objective values for problem~\eqref{MW_obj_original}. Since the optimal value of ~\eqref{MW_obj_original} is inherently upper-bounded by the transmit power and channel constraints, Algorithm~\ref{algo:AO} is guaranteed to converge to a suboptimal solution.

\section{Numerical Results}\label{simulation}
We now validate the effectiveness of PASS in multicast communications through numerical simulations. Specifically, we compare the proposed PASS framework with conventional MIMO techniques under the same simulation setup, and demonstrate the significant multicast rate improvements enabled by PASS.

\subsection{Simulation Setup}
For the configuration of PASS, the waveguides are deployed at a height of $h = 5$ m with a total length of $D_{\rm x}$. The other side length of the rectangular region is set to $D_{\rm y} = 5$ m, as illustrated in {\figurename}~{\ref{Fig1_Schematic}}. Moreover, the effective refractive index is assumed to be $n_{\mathrm{eff}} = 1.44$~\cite{10_ding2024flexible}.

For the multicast configuration, the carrier frequency is set to $f_c = 28$ GHz, and the noise power at each user is assumed to be $\sigma^2 = -90$ dBm. The users are assumed to be uniformly distributed within the service region. All numerical results are obtained by averaging over $1000$ independent random channel realizations.

\subsection{Baseline Architectures}
We compare the PASS against four benchmark MIMO technologies: \textit{conventional MIMO}, \textit{analog beamforming}, \textit{massive MIMO}, and \textit{hybrid MIMO}. Specifically, the following configurations are considered for the baselines:
\begin{itemize}
    \item[(1)] The {\bf\textit{conventional MIMO}} system is equipped with $N$ antennas in the single-waveguide scenario, and $M$ antennas in the multiple-waveguide scenario. Each antenna is connected to a dedicated RF chain, and the system employs fully digital signal processing. 
    
    \item[(2)] The {\bf\textit{analog beamforming}} system is equipped with $N$ phase shifters for each antenna, serving as a baseline method only for the single waveguide scenario. All phase shifters are connected to a single RF chain. In this system, the analog beamforming vector is derived from the conventional multicast beamformer based on the SOCP method with the constant norm constraint  of each element.
    
    \item[(3)] The {\bf\textit{massive MIMO}} system is equipped with a massive array of $MN$ antennas, each with its own RF chain, to realize fully digital signal processing. The beamformer is designed using SOCP method. Due to its large number of RF chains, this baseline represents a significantly more expensive design compared to both conventional MIMO and the proposed PASS, which requires higher hardware and energy costs to support the large number of antennas and RF chains.
    
    \item[(4)] The {\bf\textit{hybrid MIMO}} system utilizes a hybrid transceiver with $M$ RF chains, each connected to $N$ antenna elements via a network of phase shifters. It employs hybrid analog-digital signal processing. While this setup imposes the same RF cost as the proposed multiple-waveguide PASS, the phase shifter network limits its practical performance when compared to the massive MIMO system.
\end{itemize}

In simulations, all baselines are based on a half-wavelength spaced uniform linear array centered within the square region at $[D_{\rm x}/2, \ 0, \ h]^{\rm T}$, and aligned along the $y$-axis.

\subsection{Single-Waveguide Scenarios}
\begin{figure*}[!t]
  \centering
  \subfloat[Average multicast rate of PASS and fixed-location antenna systems.]{%
    \includegraphics[height=0.28\linewidth]{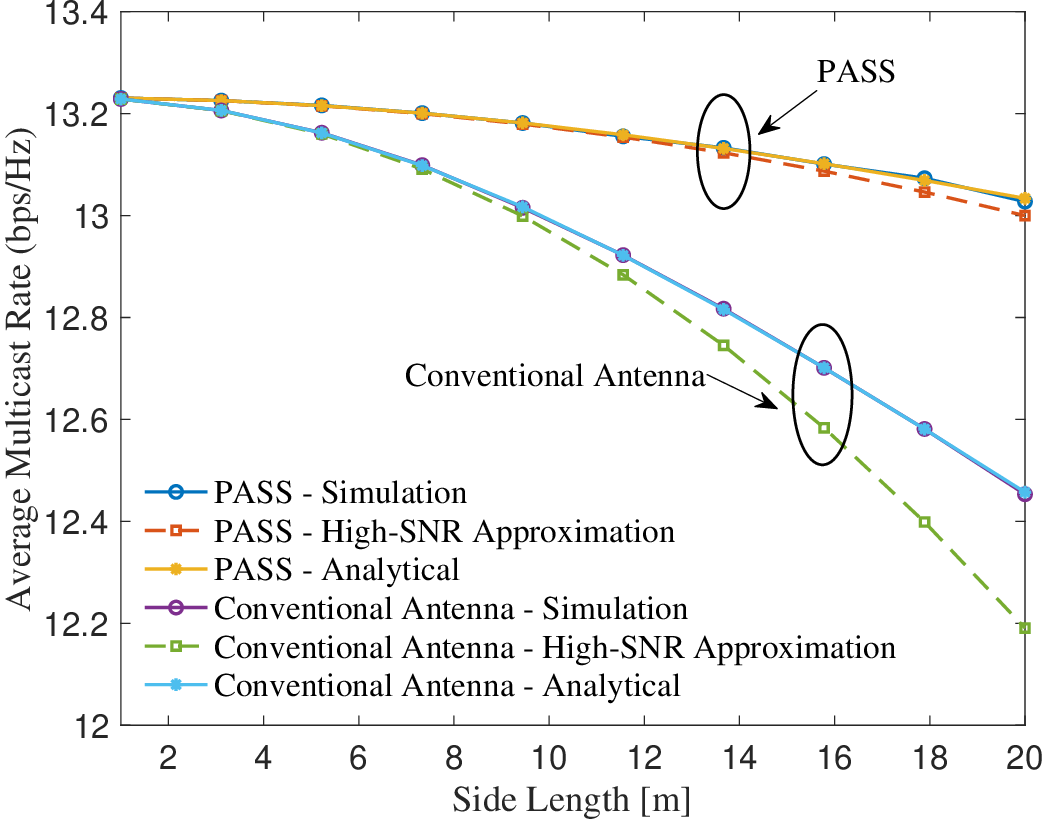}%
    \label{fig:special_rate}%
  }\hspace{0.16\textwidth}
  \subfloat[Multicast rate gain of PASS.]{%
    \includegraphics[height=0.28\linewidth]{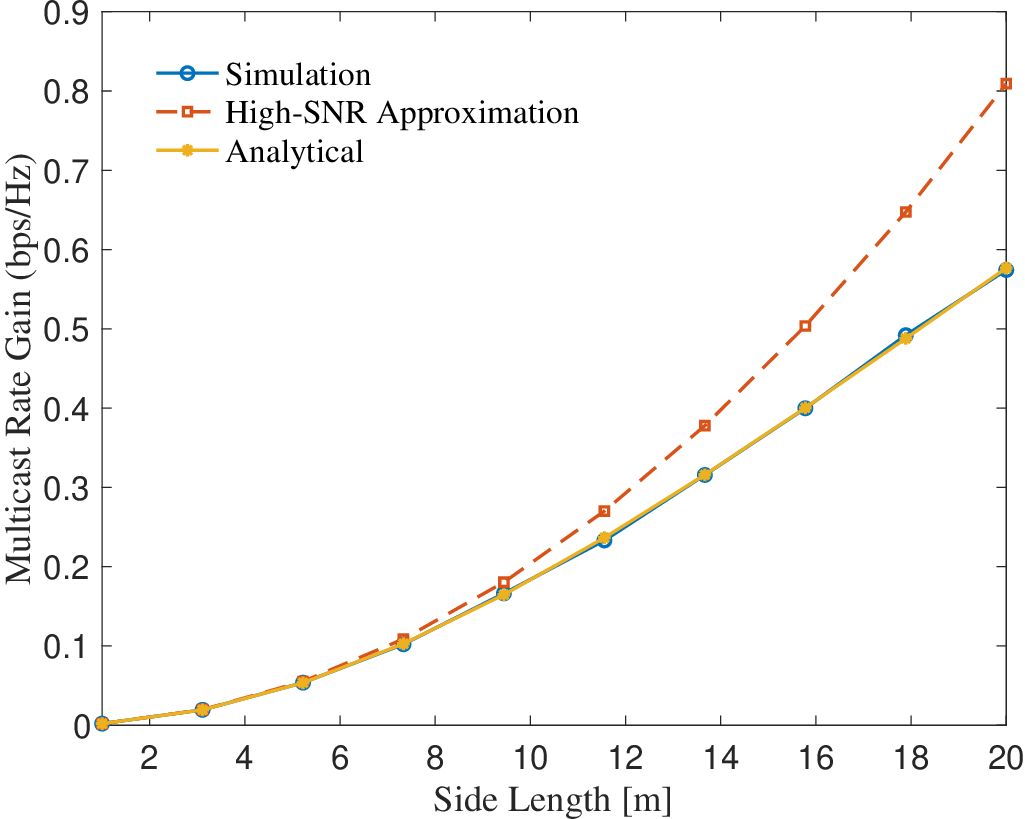}%
    \label{fig:special_gain}%
  }%
  \caption{Multicast performance versus the side length $D_{\rm{x}}$ under a single-PA deployment and a linear user distribution with $P_{\rm{t}}=10$ dBm and $K=6$. The analytical multicast rates for PASS and the conventional fixed-location antenna system are computed based on~\eqref{C_N1M12K} and~\eqref{C_conv}, respectively. The corresponding high-SNR approximations are obtained from~\eqref{PASS_appro} and~\eqref{Conv_appro}, respectively.}
  \label{fig:special_comparison}
  \vspace{-10pt}
\end{figure*}

\subsubsection{Single-PA Deployment Under Linear User Distribution}
{\figurename}~{\ref{fig:special_comparison}} compares the multicast rate performance between PASS and a conventional fixed-location antenna system under a single-PA deployment and a linear user distribution. In {\figurename}~{\ref{fig:special_rate}}, the analytical results closely match the simulation results, thereby validating the correctness of the derived expressions. For the fixed-location antenna system, the high-SNR approximation aligns well with the simulation when the side length $D_{\rm{x}}$ is small. This behavior is expected because a smaller side length results in a shorter average distance between the fixed-location antenna and the users, which creates a high-SNR environment and ensures that the approximation remains accurate. However, as $D_{\rm{x}}$ increases, the average user-to-antenna distance also grows. This leads to a lower SNR, where the high-SNR approximation becomes less accurate and gradually diverges from the simulation results. In contrast, the PASS enables the PA to reposition itself based on the user locations. This spatial flexibility allows the average distance between the PA and the users to remain approximately constant, even as $D_{\rm{x}}$ increases. As a result, the system consistently operates in a high-SNR regime, and the high-SNR approximation continues to closely match the simulation results across different side lengths. This observation highlights the superiority of PASS in its ability to dynamically adjust PAs' locations to reduce path loss and enhance multicast performance. {\figurename}~{\ref{fig:special_gain}} further illustrates the multicast rate gain achieved by PASS over the fixed-location antenna baseline. The results show that the gain increases approximately quadratically with the side length. This trend suggests that the performance advantage of PASS becomes more significant as the coverage area grows, which is consistent with the conclusion established in Remark \ref{Theorem1}.

\begin{figure}[!t]
\centering
\includegraphics[height=0.28\textwidth]{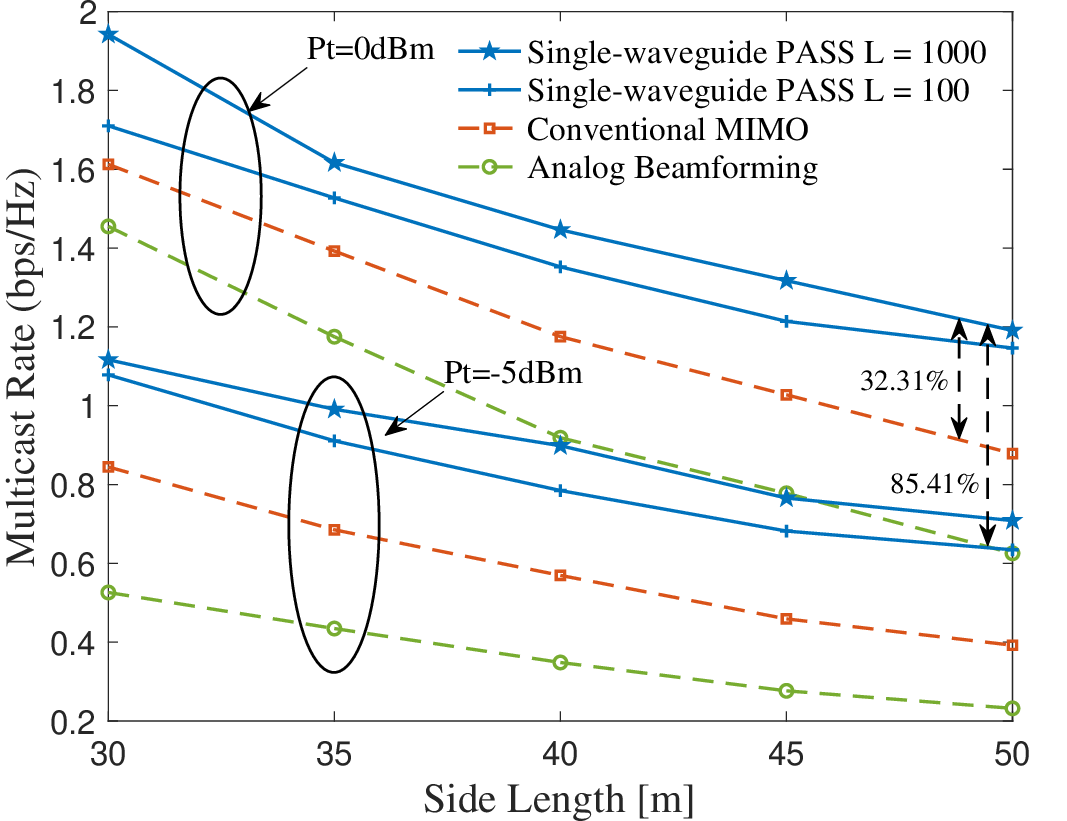}
\caption{Multicast rate versus the side length in the single-waveguide scenario with $N=6$ and $K=4$.}
\label{s_waveguide}
\vspace{-10pt}
\end{figure}

\subsubsection{Multicast Rate vs. the Side Length}
{\figurename}~{\ref{s_waveguide}} illustrates the multicast rate as a function of the side length $D_{\rm{x}}$ for various values of the one-dimensional search resolution $L$, under a single-waveguide deployment with multiple PAs. The results show that the multicast rate achieved by PASS increases as $L$ becomes larger. This trend indicates that a finer search resolution---corresponding to a greater number of candidate PA positions---offers improved flexibility in channel reconfiguration, thereby enhancing multicast performance. Furthermore, the multicast rate gain provided by PASS becomes more pronounced as the side length $D_{\rm{x}}$ increases. This observation is consistent with the results presented in {\figurename}~{\ref{fig:special_comparison}}.

Another important observation from {\figurename}~{\ref{s_waveguide}} is that PASS achieves a higher multicast rate than conventional fixed-location antenna systems across all considered $D_{\rm{x}}$ values. This includes both fully digital MIMO systems and analog beamforming architectures. It is important to note that fully digital MIMO requires significantly higher hardware complexity than PASS, as it demands a dedicated RF chain per antenna. Analog beamforming systems allow for more flexible beamformer designs by enabling continuous control over the phase shifts of the transmit signals. Despite these architectural advantages, PASS still outperforms both conventional systems in terms of multicast rate, owing to its ability to dynamically adapt antenna positions through \emph{pinching beamforming}. These results validate the effectiveness of PASS as a promising and efficient solution for future wireless networks.

\begin{figure}[!t]
\centering
\includegraphics[height=0.28\textwidth]{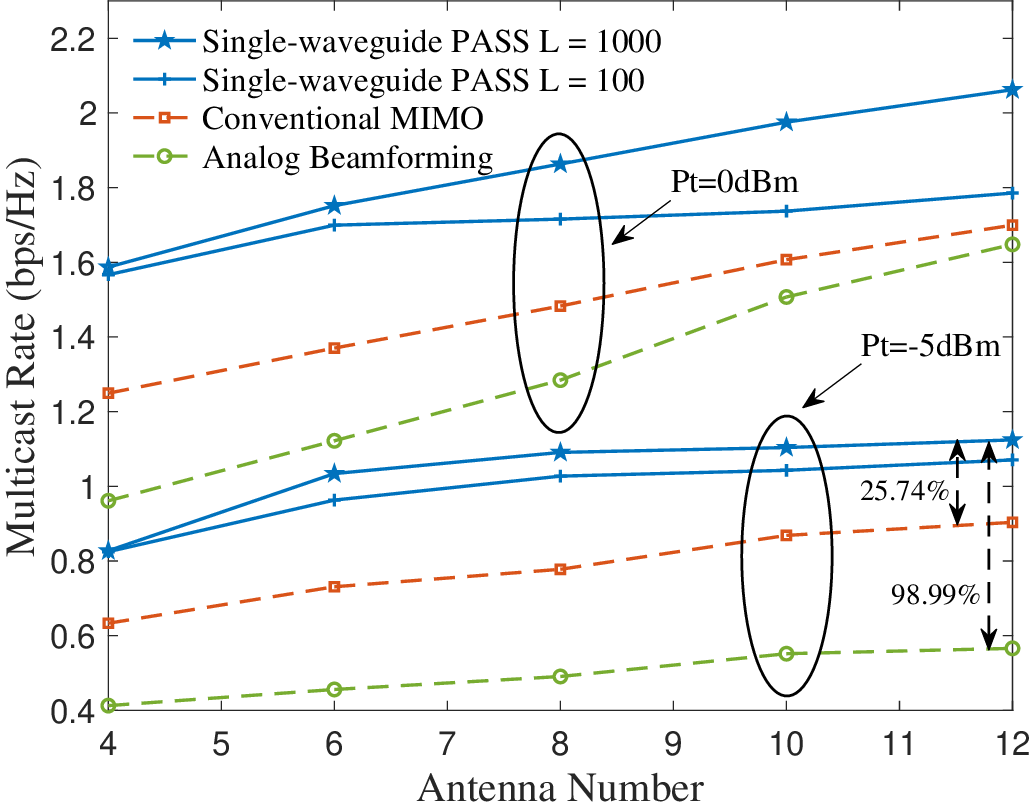}
\caption{Multicast rate versus the antenna number in the single-waveguide scenario with $K=4$ and $D_{\rm{x}}=30$ m.}
\label{s_PAnumber}
\vspace{-10pt}
\end{figure}

\subsubsection{Multicast Rate vs. the Antenna Number}
{\figurename}~{\ref{s_PAnumber}} illustrates the multicast rate as a function of the number of antennas, $N$. The results indicate that the multicast performance of PASS improves with an increasing number of PAs. This improvement is attributed to the enhanced ability of PASS to concentrate transmit energy toward the user with the worst channel condition. Consequently, the pinching beamforming gain increases, which leads to a higher multicast transmission rate. This trend is consistent with the analytical results in \cite{12_ouyang2025array}, which showed that increasing the number of PAs yields a higher array gain. In conventional MIMO systems, the multicast rate can also be improved by increasing the number of antennas. However, this requires a proportional increase in RF chains and results in higher hardware complexity and energy consumption. In contrast, PASS maintains a relatively simple architecture while achieving significant performance gains. Notably, despite the inherent simplicity of its structure, PASS still achieves higher multicast rates than conventional MIMO systems, as shown in {\figurename}~{\ref{s_PAnumber}}. This advantage is primarily due to the robustness and flexibility of the \emph{pinching beamforming}, which enables effective channel reconfiguration and improved signal alignment without the need for complex RF front-ends.

\begin{figure}[!t]
\centering
\includegraphics[height=0.28\textwidth]{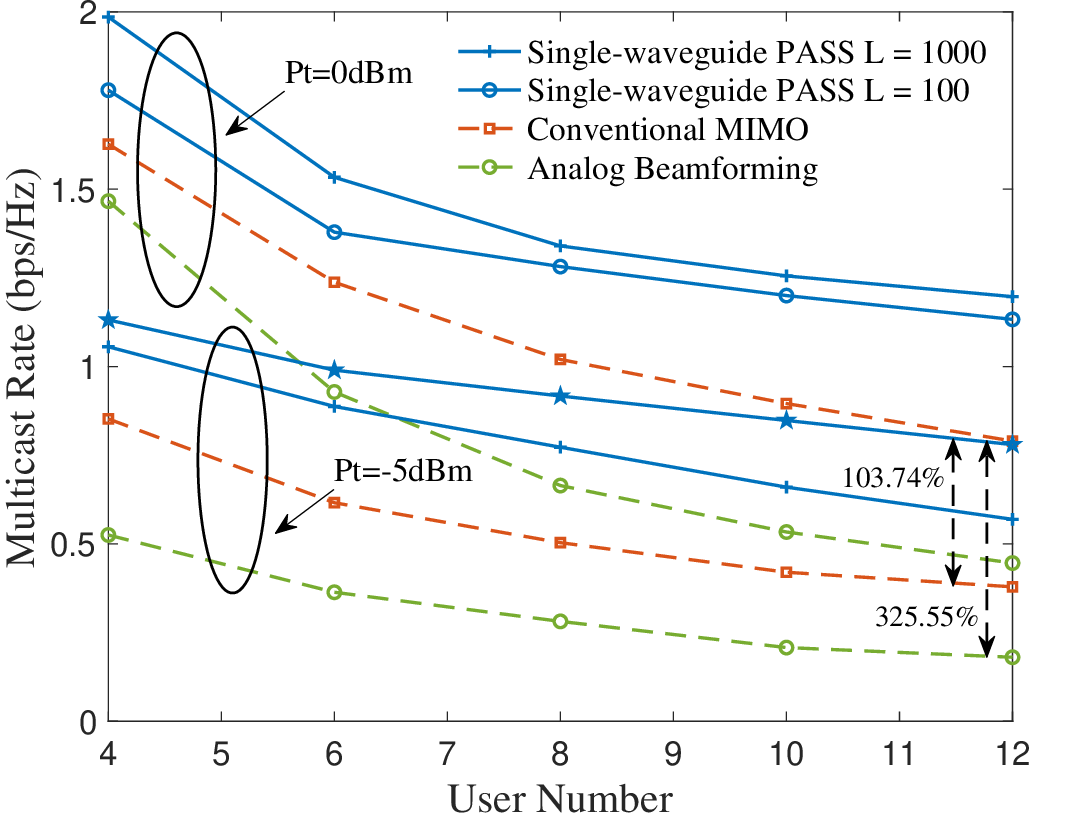}
\caption{Multicast rate versus the user number in the single-waveguide scenario with $N=8$ and $D_{\rm{x}}=30$ m.}
\label{s_UEnumber}
\vspace{-10pt}
\end{figure}

\subsubsection{Multicast Rate vs. the User Number}
{\figurename}~{\ref{s_UEnumber}} illustrates the achievable multicast rate as a function of the number of users. Across all settings, PASS outperforms both conventional MIMO and analog beamforming. As the number of users increases, the performance gap between PASS and the conventional fixed-location antenna systems becomes more pronounced. This is due to the fact that in a multicast communication scenario, the overall system performance is determined by the rate of the worst-user. With its ability to adjust PA positions, PASS ensures that each user experiences a strong LoS link. In contrast, the performance of conventional fixed-location antenna systems degrades more rapidly as the number of users increases because it is more challenging to focus the energy to the worst user for a larger number of users.

\subsection{Multiple-Waveguide Scenarios}

\begin{figure}[!t]
\centering
\includegraphics[height=0.28\textwidth]{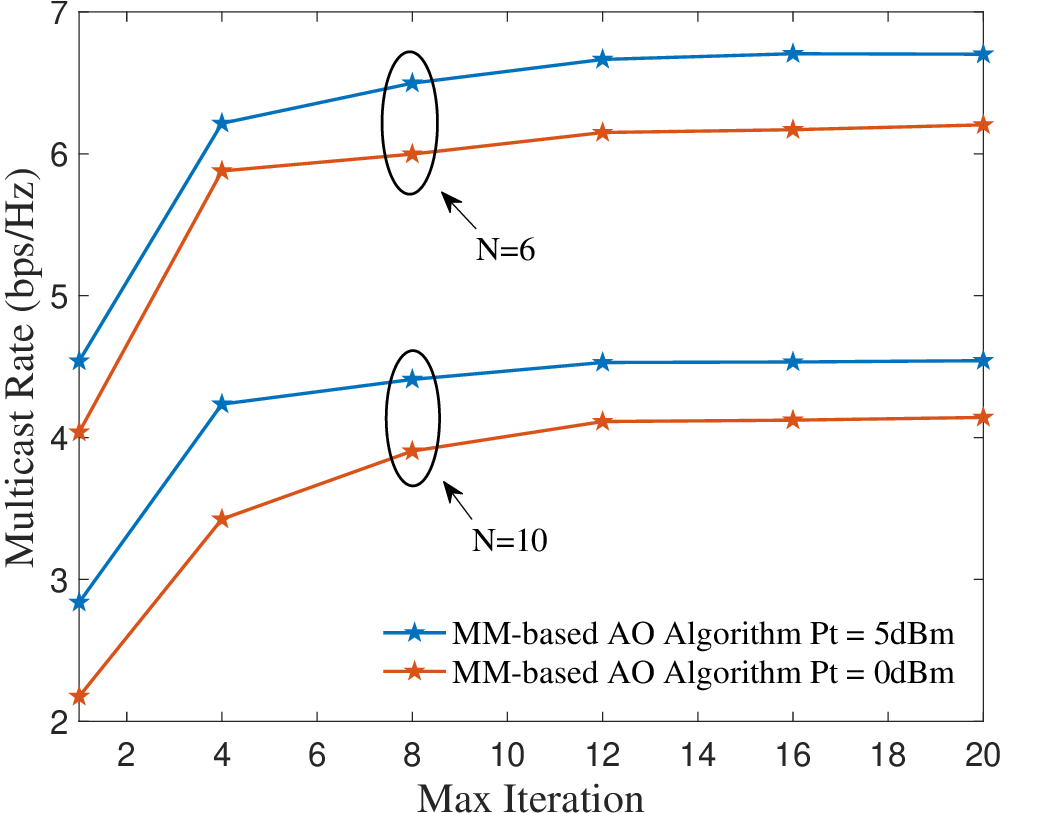}
\caption{Multicast rate versus the number of max iteration in the AO algorithm. $M=4$, $L=1000$, $K=4$, and $D_{\rm{x}}=30$~m.}
\label{Iteration}
\vspace{-10pt}
\end{figure}

\subsubsection{Convergence of the AO-Based Joint Beamforming}
We now consider the multiple-waveguide scenario, where the pinching beamformer and transmit beamformer are jointly optimized using the proposed MM-based AO algorithm (Algorithm~\ref{algo:AO}). In {\figurename}~{\ref{Iteration}}, we provide the convergence of Algorithm~\ref{algo:AO}, where both the pinching and transmit beamformers are initialized randomly. As shown in {\figurename}~{\ref{Iteration}}, the multicast rate achieved by the proposed algorithm increases rapidly with the number of iterations. This result confirms the convergence of the method and its effectiveness in joint beamforming design.

\begin{figure}[!t]
\centering
\includegraphics[height=0.28\textwidth]{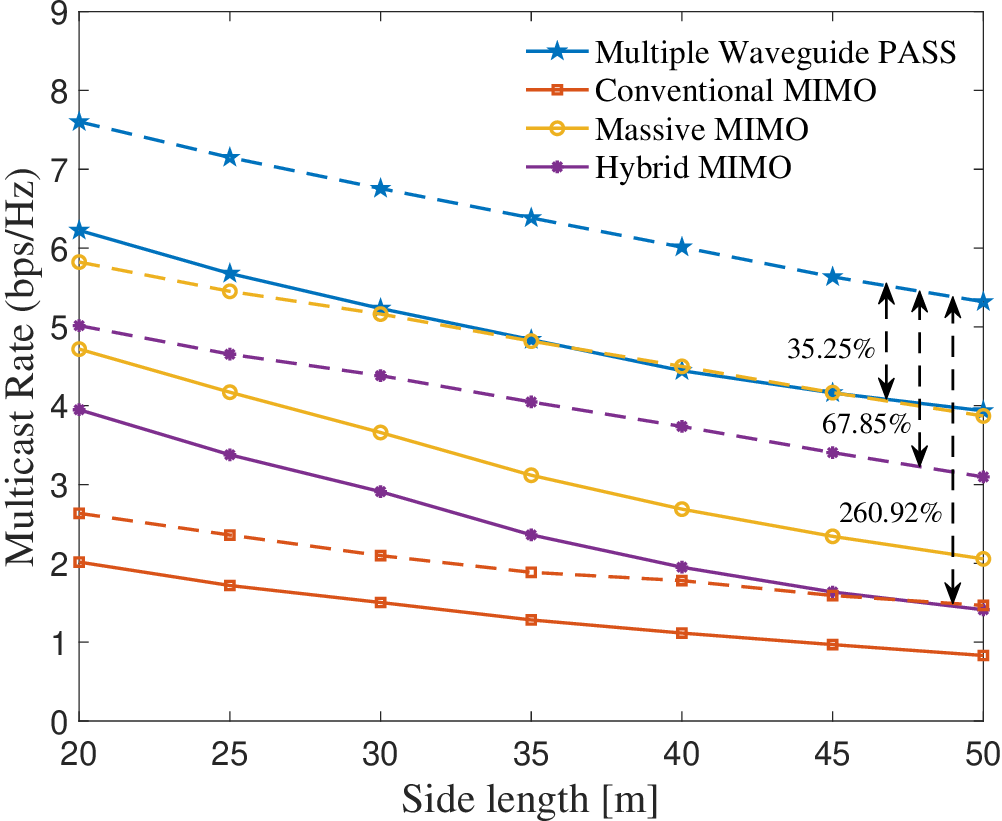}
\caption{Multicast rate versus the wavelength in the multiple-waveguide scenario under two transmit power levels: $P_{\rm t} = 0$ dBm (solid lines) and $P_{\rm t} = 5$ dBm (dashed lines). $N=12$, $M=4$, $L=1000$, and $K=4$.}
\label{m_wavelength}
\vspace{-10pt}
\end{figure}

\subsubsection{Multicast Rate vs. the Side Length}
{\figurename}~{\ref{m_wavelength}} plots the achievable multicast rate as a function of the side length. The performance of the proposed multiple-waveguide PASS is compared with that of conventional MIMO, massive MIMO, and hybrid MIMO architectures. Notably, even in the massive MIMO case---where each antenna is individually connected to a dedicated RF chain---the system still exhibits inferior multicast performance compared to PASS. This highlights the strength of the proposed approach as well as the ability of PASS to spatially reconfigure the antenna positions. These results clearly demonstrate the superiority of PASS in efficiently enhancing multicast rate performance.

\begin{figure}[!t]
\centering
\includegraphics[height=0.28\textwidth]{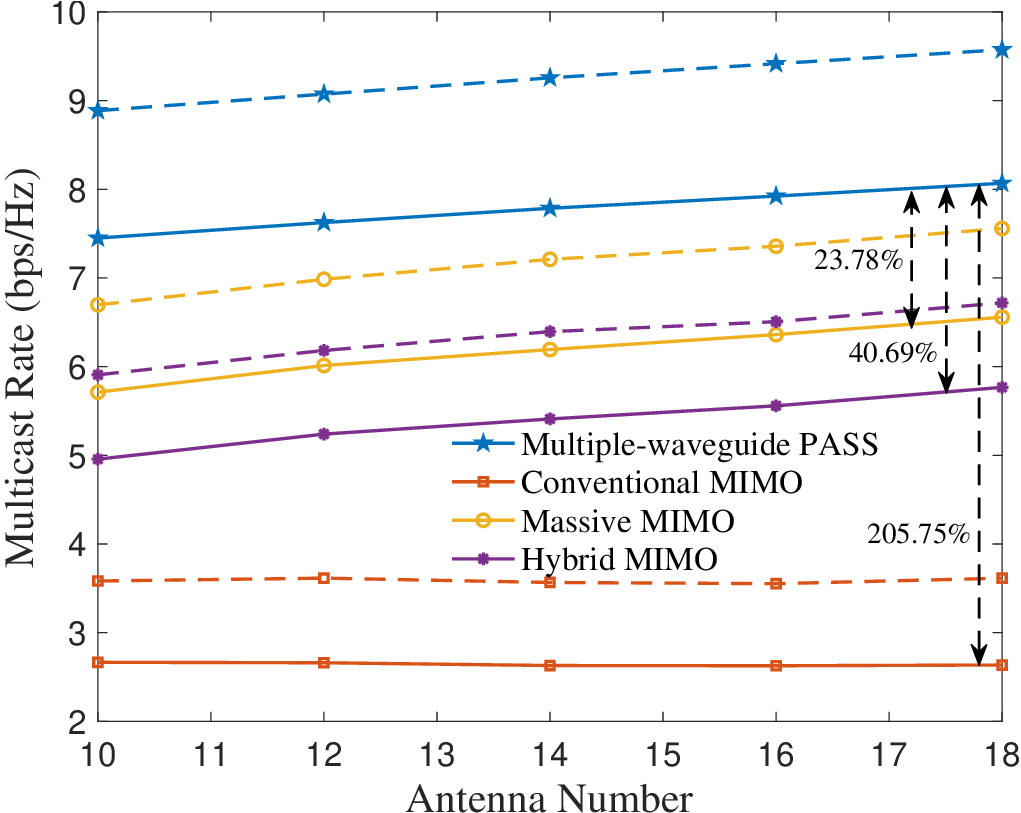}
\caption{Multicast rate versus the PA number in the multiple-waveguide scenario under two transmit power levels: $P_{\rm t} = 0$ dBm (solid lines) and $P_{\rm t} = 5$ dBm (dashed lines). $M=4$, $L=1000$, $K=4$, and $D_{\rm{x}}=30$ m.}
\label{m_PAnumber}
\vspace{-10pt}
\end{figure}

\subsubsection{Multicast Rate vs. the PA Number}
{\figurename}~{\ref{m_PAnumber}} plots the multicast rate versus the number of PAs. It can be seen from this graph that the proposed multiple-waveguide PASS outperforms conventional fixed-location antenna technologies. Besides, the multicast rate of PASS improves with an increasing number of PAs, which aligns with the trend observed in the single-waveguide scenario discussed earlier. Moreover, although the dual phase shifts induced by signal propagation both within and outside the dielectric waveguide impose certain constraints on phase control flexibility in pinching beamforming, the multicast performance growth remains favorable. As can be observed, PASS outperforms hybrid MIMO systems that employ phase shifters with infinite resolution. This observation underscores the effectiveness of the proposed joint beamforming strategy in harnessing both array gain and spatial adaptability. Therefore, we conclude that with PASS, a promising multicast performance can be achieved under simple hardware constraints.

\begin{figure}[!t]
\centering
\includegraphics[height=0.28\textwidth]{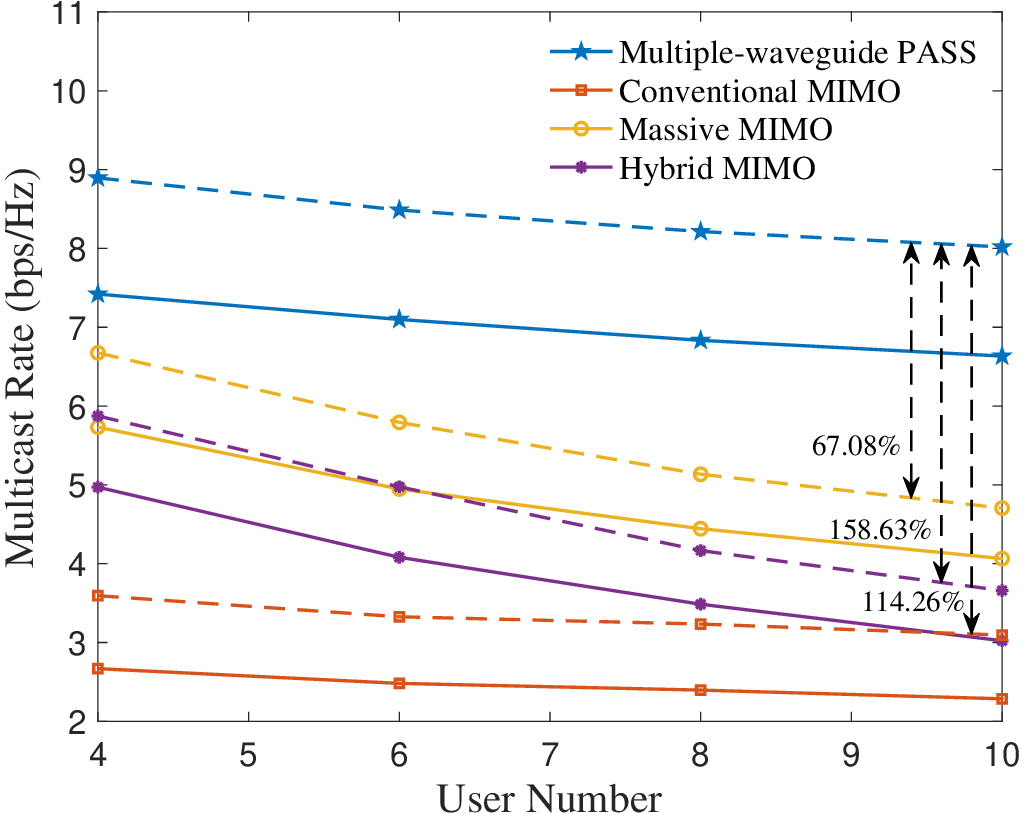}
\caption{Multicast rate versus the user number in the multiple-waveguide scenario under two transmit power levels: $P_{\rm t} = 0$ dBm (solid lines) and $P_{\rm t} = 5$ dBm (dashed lines). $M=4$, $L=1000$, $N=12$, and $D_{\rm{x}}=30$ m.}
\label{m_uenumber}
\vspace{-10pt}
\end{figure}

\subsubsection{Multicast Rate vs. the User Number}
{\figurename}~{\ref{m_uenumber}} illustrates the multicast rate as a function of the number of users. As the user number increases, the performance advantage of PASS becomes more pronounced, which is consistent with the results observed in {\figurename}~{\ref{s_UEnumber}}. In comparison to the single-waveguide scenario, the multicast rate in the multiple-waveguide PASS system declines at a significantly slower rate as the number of users grows. This is attributed to the enhanced spatial flexibility offered by the multiple-waveguide configuration. In this setup, PAs can be distributed across a two-dimensional planar region, which allows them to be positioned closer to individual users. Since the multicast rate is constrained by the user with the weakest channel condition, this spatial adaptability enables more balanced user channel gains. As a result, the system maintains strong multicast performance even in large-scale user deployments.

\section{Conclusion}
\label{conclusion}
This article investigated multicast beamforming design in PASS for multicast communications. For the single-waveguide case, we derived closed-form expressions for both the PA location and the achievable multicast rate under the assumption of a single PA and linearly distributed users. It was analytically shown that PASS achieves a higher multicast rate than conventional fixed-location antenna systems, with the performance gain becoming more pronounced in larger service regions. For scenarios involving multiple PAs and arbitrary user distributions, we proposed an element-wise AO method to design low-complexity pinching beamformers. For the multiple-waveguide case, we developed an AO-based framework to tackle the non-convex joint transmit and pinching beamforming problem. This framework combines element-wise sequential optimization for pinching beamforming with SOCP for transmit beamforming. Both analytical and numerical results demonstrated that PASS significantly improves multicast performance compared to conventional fixed-location antenna architectures--even outperforming fully digital massive MIMO systems in certain configurations. These findings suggest that PASS is a cost-effective and energy-efficient solution for enhancing physical-layer multicast in the next-generation wireless networks.

\appendices
\section{Proof for Lemma \ref{lemma1}}\label{proof-lemma1}

Based on \eqref{R_min_special}, the average multicast rate is given by
\begin{subequations}
\begin{align}
	C_{\rm PASS} & = {\mathbb{E}}\left\{R_{\rm min}(\Delta_x)\right\} \\
	& = {\mathbb{E}}\left\{\log\left(1+\frac{\eta P_{\rm t}}{D_k^2({\Delta_x})\sigma^2_k}\right)\right\}.
\end{align}
\end{subequations}
Furthermore, by denoting $A = h^2 + \hat{y}^2$ and $P_{\rm eff} = {\eta P_t}/{\sigma^2}$, the instantaneous multicast rate can be written as follows:
\begin{align}
	R_{\min}(\Delta_x) = \log_2\left(1 + \frac{P_{\rm eff}}{\Delta_x^2/4 + A}\right).
\end{align}
When ${\hat x}_{\rm min}, {\hat x}_{\rm max} \sim \mathcal{U}[\,0, \,D_{\rm x}\,] $, the probability density function (PDF) of $\Delta_x$ is given by
\begin{align}
	f_{\Delta_x}(d) = \frac{2(D_{\rm x} - d)}{D_{\rm x}^2}, \quad 0 \le d \le D_{\rm x}.
\end{align}
The average multicast rate can be calculated as follows:
\begin{align}\label{c_min}
	&C_{\rm PASS}
= \int_0^{D_{\rm x}} R_{\min}(d) \, f_{\Delta_x}(d) \, {\rm d}d\notag \\
&= \frac{2}{D_{\rm x}^2} \int_0^{D_{\rm x}} (D_{\rm x} - d)
\log_2\left(1 + \frac{P_{\rm eff}}{d^2/4 + A} \right) {\rm d}d\notag \\
&= \frac{2}{D_{\rm x}^2 \ln 2} \int_0^{D_{\rm x}} (D_{\rm x} - d)
\ln\left(1 + \frac{P_{\rm eff}}{d^2/4 + A} \right) {\rm d}d.
\end{align}
Upon using the identity
\begin{align}
	\ln\left(1 + \frac{P_{\rm eff}}{X} \right) = \ln(X + P_{\rm eff}) - \ln(X), \, X = \frac{d^2}{4} + A 
\end{align}
and defining the weighted integral
\begin{align}\label{J_C}
	J(C) = \int_0^{D_{\rm x}} (D_{\rm x} - d) \ln\left( \frac{d^2}{4} + C \right) {\rm d}d,
\end{align}
the average multicast rate \eqref{c_min} can be rewritten as follows:
\begin{align}
	C_{\rm PASS} = \frac{2}{D^2 \ln 2} ( J(A + P_{\rm eff}) - J(A) ).
\end{align}
Let $d = 2d'$, where $d'\in[0, T]$ with $T = D_{\rm x}/2$. Substituting it into~\eqref{J_C} gives
\begin{align}
	J(C) &= 2 \int_0^T (D_{\rm x} - 2d') \ln(d'^2 + C) {\rm d}d'\notag \\
&= 2D_{\rm x} \int_0^T \ln(d'^2 + C) {\rm d}d'
- 4 \int_0^T x \ln(d'^2 + C) {\rm d}d'.
\end{align}
Denoting
\begin{align}
	& I_1(C) = 2 \int_0^T \ln(d'^2 + C) {\rm d}d',\notag \\ \quad
	 & = 2\left( T \ln(T^2 + C) - 2T + 2\sqrt{C}\,\arctan\left( \frac{T}{\sqrt{C}} \right) \right)
\end{align}
and 
\begin{align}
	& I_2(C) = 4 \int_0^T d' \ln(d'^2 + C) {\rm d}d' \notag \\
	& = 2\left(
({T^2 + C}) \ln(T^2 + C)
- {T^2}
- {C} \ln C
\right),
\end{align}
the results in~\eqref{C_N1M12K} can be directly obtained.
This concludes the proof of Lemma~\ref{lemma1}.
 
 \section{Full Derivation for \eqref{PASS_appro} and \eqref{Conv_appro}}\label{proof-Theorem1}
In the high-SNR regime, we have $P_{\rm eff}\gg 1$. Therefore, \eqref{c_min} can be simplified as follows:
 \begin{align}\label{C_pass}
C_{\rm PASS}\approx \log_2 (P_{\rm eff}) \;-\;\underbrace{\int_0^{D_{\rm x}}\log_2\!\bigl(d^2/4+A\bigr)\,f_{\Delta_{x}}(d)\,{\rm d}d}_{G_{\rm PASS}}.
 \end{align}
 Using the Taylor expansion
 \begin{align}\label{taylor_approx_1}
 	\ln\bigl(d^2/4+A\bigr)\approx\ln A + \frac{d^2}{4A},
 \end{align}
 $G_{\rm PASS}$ in~\eqref{C_pass} can be approximated as follows: 
 \begin{align}\label{taylor_approx_2}
 	G_{\rm PASS}\approx \frac{\ln A}{\ln2}
+
\frac{1}{4A\ln2}{\mathbb E}\{\Delta_{x}\}
 \end{align}
 with
 \begin{align}\label{E_PASS}
 	{\mathbb E}\{\Delta_{x}\} \triangleq \frac{2}{{D_{\rm x}^2}}\int_0^{D_{\rm x}} d^2({D_{\rm x}}-d){\rm d}d =\frac{{D_{\rm x}^2}}{6}.
 \end{align} 
 Substituting~\eqref{E_PASS} into~\eqref{taylor_approx_2}, the results in~\eqref{PASS_appro} can be directly obtained.

We now consider the fixed-location antenna system. In this scenario, the users are distributed as $x_k\sim {\mathcal U}[\,0,\, D_{\rm x}\,]$. Denote $\Delta_k = |x_k - D_{\rm x}/2|\in[\,0,\,D_{\rm x}/2\,]$ as the $x$-axis distance between the user $k$ and the fixed-location antenna, then
\begin{align}
	F_{\Delta_k}(y)=\Pr\bigl\{|x_k-D_{\rm x}/2|\le y\bigr\} = \frac{2y}{D_{\rm x}},\, 0\le y\le \frac {D_{\rm x}} 2.
\end{align}
We note that $\Delta_k$ is uniform on $[\,0,\,D_{\rm x}/2\,]$ with density $f_{\Delta_k}(y)=2/D_{\rm x}$. Therefore, the maximum  $\Delta_{\rm Conv}=\max_k{\Delta_k}$ can be given by
\begin{align}
F_{\Delta_{\rm Conv}}(d) = \bigl(F_{\Delta_k}(d)\bigr)^K = \Bigl(\frac{2d}{D_{\rm x}}\Bigr)^K, \quad 0\le d\le \frac {D_{\rm x}}2,
\end{align}
and differentiating gives the PDF 
\begin{align}
	f_{\Delta_{\rm Conv}}(d) =\frac{d}{{\rm d}d}F_{\Delta_{\rm Conv}}(d) =\frac{2K}{D_{\rm x}}\Bigl(\frac{2d}{D_{\rm x}}\Bigr)^{K-1},\,\,  0\le d\le \frac {D_{\rm x}}2.
\end{align}
In the high-SNR regime, the multicast rate can be approximated as follows:
\begin{subequations}
\begin{align}
	& C_{\rm Conv}  =\int_0^{D_{\rm x}/2}\log_2\!\Bigl(1+\frac{P_{\rm eff}}{d^2+C}\Bigr)\,f_{\Delta_{\rm Conv}}(d)\,{\rm d}d\label{C_conv} \\
	& \;\approx\; \log_2 (P_{\rm eff}) \;-\; \underbrace{\int_0^{D_{\rm x}/2}\log_2(d^2+C)\,f_{\Delta_{\rm Conv}}(d)\,{\rm d}d}_{G_{\rm Conv}},
\end{align}
\end{subequations}
where
\begin{align}
	G_{\rm Conv} =\frac{1}{\ln2} \int_0^{D_{\rm x}/2}\ln(d^2+C)\;\frac{2^K\,K}{D_{\rm x}^K}\,d^{\,K-1}\,{\rm d}d.
\end{align}
According to \eqref{taylor_approx_1} and \eqref{taylor_approx_2}, \eqref{C_conv} can be further derived as follows:
\begin{align}\label{C_appro}
	C_{\rm Conv} \approx \log_2\Bigl(\frac{P_{\rm eff}}{A}\Bigr) - {\frac{1}{A\ln2}\,{\mathbb E}\{\Delta_{\rm Conv}\}},
\end{align}
where
\begin{align}\label{E_conv}
	 {\mathbb E}\{\Delta_{\rm Conv}\}&\triangleq \int_0^{D_{\rm x}/2}d^2\,f_{\Delta_{\rm Conv}}(d)\,{\rm d}d\notag \\
	 &  =\frac{2^K\,K}{D_{\rm x}^K} \int_0^{D_{\rm x}/2}d^{K+1}\,{\rm d}d\notag \\
   & =\frac{2^K\,K}{D_{\rm x}^K}\;\frac{\bigl(D_{\rm x}/2\bigr)^{K+2}}{K+2} = \frac{K}{4\,(K+2)}\,D_{\rm x}^2.
\end{align}
 Substituting~\eqref{E_conv} into~\eqref{C_appro}, the results in~\eqref{Conv_appro} can be directly obtained.

\section{Proof for Lemma \ref{lemma_surrogate}}\label{proof-lemma2}

We first rewrite the user rate expression to reveal its underlying convexity. Define auxiliary variables as $g_k \triangleq \mathbf { h}_{{\rm eff}, k}^{\rm H}\mathbf { w}$ and $r_k = \sigma_k^2 + |g_k|^2$. Then, the user rate is expressed as follows:
\begin{align}
    R_k(g_k, r_k) &= -\log_2\left(1 - \frac{|g_k|^2}{r_k}\right).
\end{align}
It can be verified that $R_k(g_k, r_k)$ is jointly convex with respect to the pair $(g_k,r_k)$. Thus, we apply the first-order approximation around a feasible point $(g_k^{(t)}, r_k^{(t)})$ to obtain a concave lower-bound surrogate function as follows:
\begin{equation}
\begin{split}
    &R_k(t_k, r_k)
    \geq R_k(g_k^{(t)}, r_k^{(t)}) + \left.\frac{\partial R_k}{\partial g_k}\right|_{g_k^{(t)}}(g_k - g_k^{(t)}) \\
    &+ \left.\frac{\partial R_k}{\partial t_k^*}\right|_{g_k^{(t),*}}(g_k^* - g_k^{(t),*})+ 
    \left.\frac{\partial R_k}{\partial r_k}\right|_{r_k^{(t)}}(r_k - r_k^{(t)}).
\end{split}
\end{equation}

Calculating the partial derivatives and simplifying, we obtain:
\begin{equation}
\begin{split}
   & R_k(g_k, r_k) \geq R_k(g_k^{(t)}, r_k^{(t)}) \\
    & + 2\Re\left\{\frac{g_k^{(t),*}(g_k - g_k^{(t)})}{\ln2\,(r_k^{(t)} - |g_k^{(t)}|^2)}\right\} - \frac{|g_k^{(t)}|^2(r_k - r_k^{(t)})}{\ln{2}r_k^{(t)}(r_k^{(t)} - |g_k^{(t)}|^2)}\\
    &= R_k\left(g_k^{(t)}, r_k^{(t)}\right) \\
    & + 2\Re\left\{\frac{g_k^{(t),*}g_k}{\ln{2}(r_k^{(t)} - |g_k^{(t)}|^2)}\right\} - \frac{|g_k^{(t)}|^2 r_k}{\ln{2} r_k^{(t)}(r_k^{(t)} - |g_k^{(t)}|^2)}.
\end{split}
\end{equation}

Substituting back $g_k = \mathbf { h}_{{\rm eff}, k}^{\rm H}\mathbf { w}$ and omitting constant terms unrelated to optimization variables, we obtain the following simplified surrogate expression:
\begin{align}
    R_k(\mathbf { w}) &\geq 2\Re\left\{a_k\mathbf { h}_{{\rm eff}, k}^{\rm H}\mathbf { w}\right\} + b_k|\mathbf { h}_{{\rm eff}, k}^{\rm H}\mathbf { w}|^2 + c_k,
\end{align}
where $a_k$, $b_k$, and $c_k$ encapsulate terms derived from the previous expansion. This concludes the proof of Lemma~\ref{lemma_surrogate}.


\begin{thebibliography}{99}

\bibitem{1_rappaport2013millimeter} C.~E.~Shannon, ``A mathematical theory of communication,'' \emph{Bell Syst. Tech. J.}, vol.~27, no.~3, pp.~379--423, Jul.~1948.

\bibitem{4_sohrabi2017hybrid} I.~F.~Akyildiz, J.~M.~Jornet, and C.~Han, ``Terahertz band: Next frontier for wireless communications,'' \emph{Phys. Commun.}, vol.~12, pp.~16--32, 2022.

\bibitem{5_gong2020toward} S.~Gong \emph{et al.}, ``Towards smart wireless communications via intelligent reflecting surfaces: A contemporary survey,'' \emph{IEEE Commun. Surveys Tuts.}, vol.~22, no.~4, pp.~2283--2314, 4th~Quart.~2020.

\bibitem{6_wu2021intelligent} Q.~Wu and R.~Zhang, ``Intelligent reflecting surface enhanced wireless network: Joint active and passive beamforming design,'' \emph{IEEE Trans. Wireless Commun.}, vol.~19, no.~4, pp.~2207--2221, Apr.~2021.

\bibitem{7_wong2020fluid} K.-K.~Wong, A.~Shojaeifard, K.-F.~Tong, and Y.~Zhang, ``Fluid antenna systems,'' \emph{IEEE Trans. Wireless Commun.}, vol.~20, no.~3, pp.~1950--1962, Mar.~2021.

\bibitem{Shojaeifard2022mimo} A. Shojaeifard, K.-K.~Wong, K.-F.~Tong, Z. Chu, A. Mourad, A. Haghighat, I. Hemadeh, N. T. Nguyen, V. Tapio, and M. Juntti, ``MIMO evolution beyond 5G through reconfigurable intelligent surfaces and fluid antenna systems,'' \emph{Proc. IEEE}, vol. 110, no. 9, pp. 1244--1265, Sep. 2022.
    
\bibitem{New2024tutorial} W. K. New, K.-K. Wong, H. Xu, C. Wang, F. R. Ghadi, J. Zhang, J. Rao, R. Murch, P. Ram\'{i}rez-Espinosa, D. Morales-Jimenez \emph{et al.}, ``A tutorial on fluid antenna system for 6G networks: Encompassing communication theory, optimization methods and hardware designs,'' \emph{IEEE Commun. Surveys Tuts.}, early access, 2024.

\bibitem{8_zhu2023movable} L.~Zhu, W.~Ma, and R.~Zhang, ``Modeling and performance analysis for movable antenna enabled wireless communications,'' \emph{IEEE Trans. Wireless Commun.}, vol.~23, no.~6, pp.~6234--6250, Jun.~2024.

\bibitem{Zhu2025tutorial} L. Zhu, W. Ma, W. Mei, Y. Zeng, Q. Wu, B. Ning, Z. Xiao, X. Shao, J. Zhang, and R. Zhang, ``A tutorial on movable antennas for wireless networks,'' \emph{IEEE Commun. Surveys Tuts.}, early access, 2025.
    
\bibitem{9_suzuki2022pinching} A.~Fukuda, H.~Yamamoto, H.~Okazaki, Y.~Suzuki, and K.~Kawai, ``Pinching antenna: Using a dielectric waveguide as an antenna,'' \emph{NTT DOCOMO Tech. J.}, vol.~23, no.~3, pp.~5--12, 2022.

\bibitem{11_yang2025pinching} Z.~Yang, N.~Wang, Y.~Sun, Z.~Ding, R.~Schober, G.~K.~Karagiannidis, \emph{et~al.}, ``Pinching antennas: Principles, applications and challenges,'' \emph{arXiv preprint arXiv:2501.10753}, 2025.

\bibitem{28_yuanwei2025architecture} Y.~Liu, Z.~Wang, X.~Mu, C.~Ouyang, X.~Xu, and Z.~Ding, ``Pinching-antenna systems (PASS): Architecture designs, opportunities, and outlook,'' \emph{arXiv preprint arXiv:2501.18409}, 2025.

\bibitem{Wong2021vision} K.-K. Wong, K.-F.~Tong, Z. Chu, and Y. Zhang, ``A vision to smart radio environment: Surface wave communication superhighways,'' \emph{IEEE Wireless Commun.}, vol. 28, no. 1, pp. 112--119, Feb. 2021.

\bibitem{Liu2024} H. Liu, W. K. New, H. Xu, Z. Chu, K.-F. Tong, K.-K. Wong, and Y. Zhang, ``Path loss and surface impedance models for surface wave-assisted wireless communication system,'' \emph{IEEE Access}, vol. 12, pp. 125786--125799, 2024.
        
\bibitem{Chu2024} Z. Chu, K.-F. Tong, K.-K. Wong, C.-B. Chae, and C. H. Chan, ``On propagation characteristics of reconfigurable surface wave platform: Simulation and experimental verification,'' \emph{IEEE Access}, vol. 12, pp. 168744--168754, 2024. 

\bibitem{10_ding2024flexible} Z.~Ding, R.~Schober, and H.~V.~Poor, ``Flexible-antenna systems: A pinching-antenna perspective,'' \emph{IEEE Trans. Commun.}, early access, 2025.

\bibitem{31_Tyrovolas2025performance} D.~Tyrovolas, S.~A.~Tegos, P.~D.~Diamantoulakis, S.~Ioannidis, C.~K.~Liaskos, and G.~K.~Karagiannidis, ``Performance analysis of pinching-antenna systems,'' \emph{IEEE Trans. Green Commun. Netw.}, early access, 2025.

\bibitem{12_ouyang2025array} C.~Ouyang, Z.~Wang, Y.~Liu, and Z.~Ding, ``Array gain for pinching-antenna systems (PASS),'' \emph{IEEE Commun. Lett.}, early access, 2025.

\bibitem{28_Guizhou2025channelesti} G.~Zhou, V.~Papanikolaou, Z.~Ding, and R.~Schober, ``Channel estimation for mmWave pinching-antenna systems,'' \emph{arXiv preprint arXiv:2504.09317}, 2025.

\bibitem{27_Jianxiao2025channelesti} J.~Xiao, J.~Wang, and Y.~Liu, ``Channel estimation for pinching-antenna systems (PASS),'' \emph{arXiv preprint arXiv:2503.13268}, 2025.

\bibitem{13_tegos2024minimum} S.~A.~Tegos, P.~D.~Diamantoulakis, Z.~Ding, and G.~K.~Karagiannidis, ``Minimum data rate maximization for uplink pinching-antenna systems,'' \emph{IEEE Wireless Commun. Lett.}, vol.~14, no.~5, pp.~1516--1520, May 2025.

\bibitem{25_Tianwei2025uplink} T.~Hou, Y.~Liu, and A.~Nallanathan, ``On the performance of uplink pinching antenna systems (PASS),'' \emph{arXiv preprint arXiv:2502.12365}, 2025.

\bibitem{15_xu2024rate} Y.~Xu, Z.~Ding, and G.~K.~Karagiannidis, ``Rate maximization for downlink pinching-antenna systems,'' \emph{IEEE Wireless Commun. Lett.}, vol.~14, no.~5, pp.~1431--1435, May 2025.

\bibitem{26_zhaolin2025Modeling} Z.~Wang, C.~Ouyang, X.~Mu, Y.~Liu, and Z.~Ding, ``Modeling and beamforming optimization for pinching-antenna systems,'' \emph{arXiv preprint arXiv:2502.05917}, 2025.

\bibitem{25_bereyhi2025MIMOPASS} A.~Bereyhi, S.~Asaad, C.~Ouyang, Z.~Ding, and H.~V.~Poor, ``Downlink beamforming with pinching-antenna assisted MIMO systems,'' in \emph{Proc. IEEE ICC Workshops}, 2025.

\bibitem{17_guo2025deep} J.~Guo, Y.~Liu, and A.~Nallanathan, ``GPASS: Deep learning for beamforming in pinching-antenna systems (PASS),'' \emph{arXiv preprint arXiv:2502.01438}, 2025.

\bibitem{24_Xiaoxia2025Ioint} X.~Xu, X.~Mu, Y.~Liu, and A.~Nallanathan, ``Joint transmit and pinching beamforming for pinching antenna systems (PASS): Optimization-based or learning-based?,'' \emph{arXiv preprint arXiv:2502.08637}, 2025.

\bibitem{14_wang2024antenna} K.~Wang, Z.~Ding, and R.~Schober, ``Antenna activation for NOMA assisted pinching-antenna systems,'' \emph{IEEE Wireless Commun. Lett.}, vol.~14, no.~5, pp.~1526--1530, May 2025.

\bibitem{29_zhao2025WDMA} J.~Zhao, X.~Mu, K.~Cai, Y.~Zhu, and Y.~Liu, ``Waveguide division multiple access for pinching-antenna systems (PASS),'' \emph{arXiv preprint arXiv:2502.17781}, 2025.

\bibitem{30_zhangzheng2025ISAC} Z.~Zhang, Y.~Liu, B.~He, and J.~Chen, ``Integrated sensing and communications for pinching-antenna systems (PASS),'' \emph{arXiv preprint arXiv:2504.07709}, 2025.

\bibitem{Unimodality} E.~M.~J.~Bertin, I.~Cuculescu, and R.~Theodorescu, \emph{Unimodality of Probability Measures}. Springer, Berlin, Germany, 2013.

\bibitem{Chebyshev} F.~Plastria and E.~Carrizosa, ``Minmax-distance approximation and separation problems: geometrical properties,'' \emph{Mathematical Programming}, vol.~133, no.~1--2, pp.~173--195, 2012.

\bibitem{Boyd} S.~Boyd and L.~Vandenberghe, \emph{Convex Optimization}. Cambridge Univ. Press, Cambridge, U.K., 2004.

\bibitem{21_gorski2007biconvex} J.~Gorski, F.~Pfeuffer, and K.~Klamroth, ``Biconvex sets and optimization with biconvex functions: A survey and extensions,'' \emph{Math. Oper. Res.}, vol.~66, no.~3, pp.~373--407, Aug.~2007.

\bibitem{22_CVX2018} M.~Grant and S.~Boyd, ``CVX: MATLAB software for disciplined convex programming,'' \emph{Version 2.1. [Online]. Available: http://cvxr.com/cvx}, Dec.~2018.

\bibitem{23_Mosek2018} ``The MOSEK optimization toolbox for MATLAB manual,'' \emph{Version 7.1 (revision 28). [Online]. Available: http://mosek.com}, accessed on: Mar.~20, 2015.

\end{thebibliography}
\end{document}